\documentclass[nopreprint,twocolumn,a4paper]{jasatex}
\usepackage{graphicx}% Include figure files
\begin{document}
%%%%%%%%%%%%%%%%%%%%%%%%%%%%%%%
\setlength{\topmargin}{-1.0cm}
%%%%%%%%%%%%%%%%%%%%%%%%%%%%%%%

\title[Numerical study on an air-reed instrument with LES
%Running Title
]{Numerical study on 
sound vibration of an air-reed instrument
with compressible LES}

% repeat the \author .. \affiliation  etc. as needed
% \email, \thanks, \homepage, \altaffiliation all apply to the current
% author. Explanatory text should go in the []'s, actual e-mail
% address or url should go in the {}'s for \email and \homepage.
% Please use the appropriate macro foreach each type of information

% \affiliation command applies to all authors since the last
% \affiliation command. The \affiliation command should follow the
% other information
% \affiliation can be followed by \email, \homepage, \thanks as well.
\author{Masataka Miyamoto}
\author{Yasunori Ito}
\author{Kin'ya  Takahashi}
\email[Kin'ya Takahashi email:]{takahasi@mse.kyutech.ac.jp}

%\homepage[]{Your web page}
%\thanks{}
%\altaffiliation{}
\affiliation{The Physics Laboratories, Kyushu Institute of Technology,
Kawazu 680-4, Iizuka 820-8502, Japan%
 %Author's affiliation
}
\author{Toshiya Takami}
\author{Taizo Kobayashi}
\author{Akira Nishida}
\author{Mutsumi Aoyagi} 
\affiliation{Research Institute for Information Technology, Kyushu
University,
6-10-1 Hakozaki, Higashi-ku, Fukuoka 812-8581, Japan}
\date{\today}

\begin{abstract}
Acoustic mechanics of air-reed instruments is investigated numerically
with compressible Large-eddy simulation (LES). Taking a two dimensional 
air-reed instrument model, we have succeeded in reproducing sound 
oscillations excited in the resonator and have studied the
characteristic feature of air-reed instruments, i.e., the relation of the
sound frequency with the jet velocity described by the
semi-empirical theory developed by Cremer \& Ising, Coltman and 
other authors based on experimental results.  
\end{abstract}

\pacs{43.75.Qr,43.75.Np,43.75.Ef,43.75.-z,43.28.Ra
%43.75.Qr 	Flutes and similar wind instruments 
%43.75.Np 	Pipe organs
%43.75.Ef 	Woodwinds
%43.75.-z 	Music and musical instruments
%43.50.Nm 	Aerodynamic and jet noise (see also 43.28.Ra)
%43.50.-x 	Noise: its effects and control
%43.28.-g 	Aeroacoustics and atmospheric sound
%43.28.Kt 	Aerothermoacoustics and combustion acoustics 
%43.28.Py 	Interaction of fluid motion and sound, Doppler effect, and sound in flow ducts
%43.28.Ra 	Generation of sound by fluid flow, aerodynamic sound and turbulence 
%47.85.Gj 	Aerodynamics
%47.27.ep 	Large-eddy simulations
% insert suggested PACS numbers
}

%\maketitle must follow title, authors, abstract, \pacs, and \keywords
\maketitle

% body of paper here - Use proper section commands
% References should be done using the \cite, \ref, and \label commands
\section{Introduction}
% Put \label in argument of \section for cross-referencing
%\section{\label{}}
%\subsection{}
%\subsubsection{}

Elucidation of acoustical mechanism of air-reed instruments is 
a long standing problem in the field of musical acoustics and is still
not understood completely\cite{PhysMI,Hirschberg}.  
The sound source of those instruments is a sort of aerodynamic sound source,
so-called edge tone, which is generated by an oscillating jet flow collided
with an edge\cite{Brown,Howe2}. Actually, the oscillating jet flow,
which emanates from a flue, passes through an open mouth and collides 
with the edge, plays the role of sound source. 
The major difficulty of analysing 
the acoustic mechanism of air-reed instruments is in strong and complex 
interactions between the air flow dynamics acting as the sound source and a
sound field excited in a resonator by it\cite{PhysMI,Hirschberg,Howe2}. 
Indeed, the resonance of a pipe induces the sound field exceeding 
$140{\rm dB}$ in it so that the strong sound field must affect the motion 
of the sound source jet, if the synchronisation between the jet 
oscillation and the sound field is well sustained. In order to elucidate this
process in detail, we need to study the dynamics of the jet flow in
terms of fluid dynamics as well as that of the sound field in terms of
acoustics at the same time taking into account the complex interaction between
them. However the method necessary to attack this problem has not been
established yet.   

In the long history of studying the edge tone, some
phenomenological formulae that describe the relation of the 
edge tone frequency with the jet velocity have been proposed so far.
A very useful formula was introduced based on the experimental results 
by Brown in 1937\cite{Brown}: it indicates that the frequency is
proportional to the jet velocity. 
In the field of the musical acoustic, the phenomenological theory that
describes the behavior of instrument driven by an air jet has been
developed since 1960's\cite{PhysMI}. The mechanism that the air jet
drives the resonator was studied by several authors experimentally and 
theoretically, chief among pioneers being Cremer and Ising, and
Coltman\cite{Cremer,Coltman,Coltman2,Coltman3,Coltman4,Flethcer_Thwaites1,Flethcer_Thwaites2,Flethcer_Thwaites3,Flethcer_Thwaites4,Fletcher,Elder,Yoshikawa}. 
It was found that two type of driving mechanism, volume-flow mechanism and 
momentum-drive mechanism, simultaneously work in jet-resonator
interaction, though the volume-flow mechanism dominates in a lower range
of the jet velocity\cite{PhysMI,Cremer,Coltman2,Coltman3,Fletcher,Elder,Yoshikawa}. 
The theory that describes the behavior of the resonator driven by the
air jet has also developed with help of the equivalent
electrical circuit theory. As a result, the difference in sounding
between the pure edge tone and the air-reed instrument is clarified in
relation of the sound frequency to the jet velocity. 

However, the theory developed in the field of musical acoustics includes
many conceptual approximations and is far from rigorous.  
A rigorous theory must be framed based upon the Navier-Stokes equations of
fluid dynamics considering the role of vorticity field of 
a real fluid as a sound source in the complex geometry of the instrument. 
The aerodynamic sound source was first formulated by Lighthill 
in 1952\cite{Lighthill}. 
Lighthill introduced an inhomogeneous wave equation whose inhomogeneous
term behaves as a quadrupole source term, so-called Lighthill's acoustic
analogy. Powell and Howe followed Lighthill's work and introduced a
very important notion that the major part of Lighthill's source term
comes from unsteady motion of vortices, namely vortex
sound\cite{Powell,Howe}. 
Further, Howe discussed the acoustic mechanism of air-reed
instruments in terms of the vortex sound theory\cite{Howe2,Howe}, 
which has since been followed in some detail by Hirschberg et al
\cite{Hirschberg,Hirschberg_group,Hirschberg_group2,Hirschberg_group3,
Hirschberg_group4,Hirschberg_group5,Hirschberg_group6}. 
Nevertheless, the detail mechanism of sound generation by the air jet and
of jet-resonator interaction is still an unsolved problem due to complex
behavior of fluid in the complex geometry of the instrument, though
great efforts have been made on this problem and are
continued\cite{Hirschberg_group6,Yoshikawa2,kuhnelt,Yagawa_group,
Bamberger,Lauroa,Delgosha,Braasch,Auberlechner}.   

The final goal of our study is to analyze the interaction of the jet flow with 
the sound field excited in the resonator and to explain in terms of the
vortex sound theory the acoustic mechanism which excites and sustains a
sound field in the resonator of air-reed instruments\cite{Howe2}. 
For the first step, we numerically study sounding vibrations 
of a small air-reed
instrument by using a two-dimensional model in this paper. 
Comparing with the theory developed by Cremer \& Ising, Coltman and
other authors, we will discuss to what extent our model reproduces 
characteristic features of sound vibrations of air-reed instruments. 

For the numerical study on aero-acoustic problems, e.g., noises caused
by aircraft and high-speed trains, we normally use a 
hybrid method, in which the sound field can be separated from 
the turbulence flow acting as sound sources and they are calculated by
different schemes, each of which is suitable for the calculation of individual
dynamics.  However, it is necessary for the calculation of air-reed
instruments to solve the turbulence flow and the sound field
simultaneously, because of strong nonlinear interaction between 
the flow and the sound. Then, we use a compressible
LES (Large-Eddy Simulation) instead of the hybrid method\cite{LES}. 
The reason we adopt LES is that LES is very
stable for a long term calculation, though it somewhat sacrifices accuracy. 

The organization of the present paper is as follows. 
In section 2, we review previous studies relative to the sound production of
air-reed instruments. We briefly explain Ligthhill's theory and Brown's
work on the edge tone. Further, we explain the phenomenological theory of
air-reed instruments developed by Cremer \& Ising, Coltman and other 
researches. 
In section 3, we introduce our model instrument, a small two-dimensional
air-reed instrument with a closed end, and explain the environment of
numerical calculation. 
In section 4, we show the results of the numerical analysis.
First we explain results of stable oscillation with an optimal choice of jet
velocity. The relation of time evolution of acoustic pressure observed 
in the pipe with that of vorticity of the jet flow is investigated taking the 
correlation between them. We also show special distributions of
characteristic dynamical-variables observed at a certain time
in a near field of the instrument: air density, velocity, vorticity and
Lighthill's sound source are considered.       
Next, we study frequency change of acoustic wave excited in the pipe 
with increase of the jet velocity. We compare our numerical results with
the prediction given by the phenomenological theory of
air-reed instruments introduced in section 2.  
Section 5 is devoted to a summary and discussion. 
  
\section{Theory relative to sound production of air-reed instruments}
\subsection{Lighthill's Theory}
The sound generated by turbulence is usually called aerodynamic
sound, which is a very small byproduct of the motion of unsteady flows
of high Reynolds number. 
The source of aerodynamic sound was first given the exact form 
by Lighthill\cite{Lighthill}. 
Lighthill transformed exactly the set of fundamental equations, 
Navier-Stokes and continuity equations, to  an inhomogeneous wave
equation whose inhomogeneous term plays the role of the source:
\begin{eqnarray}
\label{eq:Lighthill}
\biggl(\frac{\partial^2 }{\partial t^2}  -c^2 \nabla^2 \biggr)
(\rho -\rho_0) = \frac{\partial^2 T_{ij}}{\partial x_i \partial x_j},
\end{eqnarray}
where the tensor $T_{ij}$ is called Lighthill's tensor and is defined by
\begin{eqnarray}
\label{eq:Lighthilltensor}
T_{ij}=\rho v_i v_j +((p-p_0)-c^2(\rho-\rho_0))\delta_{ij} +\sigma_{ij}.
\end{eqnarray}
Here, $c$ denotes the speed of sound, $p$ the air pressure with the
average $p_0$, $\rho$ the air density with the average $\rho_0$, and 
$\sigma_{ij}$ the viscous stress tensor. 
The sound generated by the localized turbulence is regarded as that wave 
propagating in a stationary acoustic medium, which is generated 
by the quadrupole source distribution
given by the inhomogeneous term in RHS of eq.(\ref{eq:Lighthill}). This
is Lighthill's acoustic analogy. 

Since the dissipation by $\sigma_{ij}$ can be ignored for a high Reynolds
number and adiabaticity is well held as  
$(p-p_0)-c^2(\rho-\rho_0)=0$, then 
the first term of eq.(\ref{eq:Lighthilltensor}), $\rho v_i v_j$, becomes
the major term of the source. 
Further, particle velocities of the sound are usually sufficiently small
comparing with those of the real flow and so the source term is well
approximated by that obtained from incompressible fluid with      
$\rho=\rho_0$ and ${\rm div}~ v=0$.
For two dimensional (2D) fluid, it is reduced into 
\begin{eqnarray}
\label{eq:2DLighthill}
\frac{\partial^2 T_{ij}}{\partial x_i \partial x_j}
\sim - 2\rho_0 \biggl(\frac{\partial v_1 }{\partial x_1 }
\frac{\partial v_2}{\partial x_2} - 
\frac{\partial v_2}{\partial x_1 }
\frac{\partial v_1}{\partial x_2} \biggr).
\end{eqnarray}
In calculation of Lighthill's source in section \ref{IV}, we will use 
this formula.

\subsection{Edge tone}
\begin{figure}[bht]
\begin{center}
%\scalebox{0.8}{\includegraphics{jet.eps}}
\includegraphics[scale=0.3]{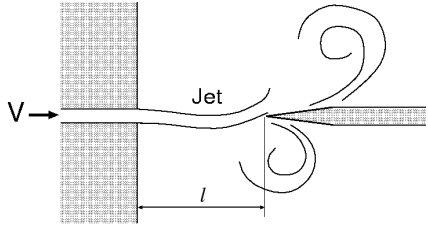}
\end{center}
\caption{Edge tone}
\label{fig-edge}
\end{figure}

As shown in Fig.\ref{fig-edge}, edge tone is an aerodynamic sound generated by 
the unsteady but mostly periodical oscillation of the jet emanated
from the flue and collied with the edge. The edge tone is the sound 
source of air-reed instruments\cite{PhysMI,Hirschberg,Howe2}. 
Although the detail mechanism of the production of edge tone has not 
been completely understood yet, its characteristic features have been
well captured by semi-empirical equations introduced based on
experimental results. One of those equations was introduced by Brown to
predict the frequency of edge tone\cite{Brown}:  
\begin{eqnarray}
\label{eq:edge}
\nu=0.466j(100V-40)(1/(100 l )-0.07), 
\end{eqnarray}
where $V$ denotes the speed of jet and $l$ is the distance between the
flue and the edge. The number $j$ is taken as
$j=1.0, 2.3, 3.8, 5.4$. For $j=1$, it gives the fundamental frequency and
others denote overtones. With increase of $V$, the fundamental
oscillation is excited and its frequency increases in proportion to
$V$. But it jumps to one of overtones, if $V$ exceeds a threshold value, 
and it jumps successively from one to other with increase of $V$. The
transitions are hysteretic so that the threshold values of $V$ of
the downward process are usually different from those of the upward.   

\subsection{Phenomenological theory of air-reed instruments}
The driving mechanism of air-reed instruments was first studied by
Cremer \& Ising, and Coltman as pioneer
works\cite{Cremer,Coltman,Coltman2,Coltman3,Coltman4}, 
which have been followed
by many authors\cite{Fletcher,Elder,Yoshikawa}. 
Those studies have made differences in driving mechanism clear 
between the edge tone and the air-reed instrument with a resonator.     
It is turned out that there are two ways in which a pulsating jet drives
a pipe, momentum drive and volume flow drive\cite{PhysMI}. 
In the momentum drive, the jet is brought to rest by mixing and dissipation, 
and thereby generates an acoustic pressure which acts on an acoustic 
flow in the resonator. 
In the volume flow drive, the jet essentially contributes a volume flow which
acts on the acoustic pressure at the open mouth, which is not
zero because of the end correction. 
In the following, we mention the outline of driving mechanism and
regenerative excitation mechanism of air-reed instruments according to
the text book by Fletcher and Rossing\cite{PhysMI}. 

\subsubsection{Driving mechanism} \label{IIC1}
\begin{figure}[bht]
\begin{center}
%\scalebox{0.8}{\includegraphics{jet.eps}}
\includegraphics[scale=0.55]{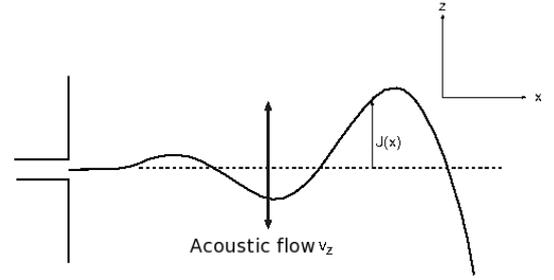}
\end{center}
\caption{Jet oscillation}
\label{fig-jet}
\end{figure}

The jet which emanates from a flue is disturbed by a uniform
transverse acoustic-flow $v_z\exp(i\omega t)$ and forms an oscillating wave
which propagates with a velocity $u$ and grows exponentially with a growth
parameter $\mu$.    
The wave of jet propagating in $x$-direction is well approximated by the
semi-empirical equation (also see Fig.\ref{fig-jet}):  
\begin{eqnarray}
J(x)= - i \frac{v_z}{\omega}\bigl\{\exp(i\omega t) -\cosh(\mu x)
\exp(i\omega(t-x/u)) \bigr\}.\nonumber\\
\label{eq:Jx}
\end{eqnarray}
In the range $kd\gg 1$, where $d$ stands for the height of
flue aperture, we can use approximations $u\sim V/2$ and $\mu\sim k$. 
Since they are available in a quite wide range, we always assume
those approximations in the following.
In this formula, the existence of edge, which plays the important
role in formation of the jet oscillation, is ignored. 
Thus it should be considered
that eq.(\ref{eq:Jx}) only gives the lowest order approximation of the jet 
motion.  

\begin{figure}[bht]
\begin{center}
%\scalebox{0.8}{\includegraphics{jet.eps}}
\includegraphics[scale=0.5]{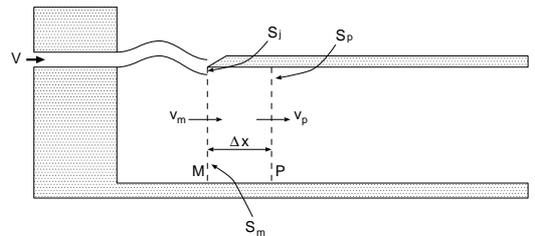}
\end{center}
\caption{Pipe driven by jet}
\label{fig-opptrecorder}
\end{figure}

Taking into account the jet oscillation given by eq.(\ref{eq:Jx}), 
it is possible to make phenomenological formulation of that driving
mechanism through which the jet acts on the pipe. 
As shown by Fig.\ref{fig-opptrecorder}, after the jet of a steady velocity 
$V$ reaches the edge
of the pipe,  a fraction of its cross section $S_j$ enters the pipe.  
It is assumed that the fraction of jet blends with an acoustic flow in
the volume between the cross sections M and P with a distance 
$\Delta x$ so that only acoustic motions survive in the other part of
pipe beyond the cross section P. 
Under this assumption, the acoustic volume flow
$U_p$ at the cross section P is expressed by 
\begin{eqnarray}
\label{eq:Up}
U_p=\frac{\rho V^2 S_j}{S_p(Z_p+Z_m)}+\frac{i\rho \omega \Delta L V
 S_j}{S_p(Z_p +Z_m)},
\end{eqnarray}
where $Z_m=i\rho \omega \Delta L/S_p$ denotes the impedance at a frequency
$\omega$ looking out from the plane M, $\Delta L$ the end correction at
the open mouth, $Z_p$ the pipe impedance evaluated at the plane M and 
$S_j$ the area of the cross section of the pipe. 

Two term in RHS of eq.(\ref{eq:Up}) indicate two different drive
mechanisms. The first constitutes the momentum drive and the second is
the volume flow drive. In the situation $\omega \Delta L > V$, the second
term is larger than the first so that the volume-flow mechanism
dominates the momentum-drive mechanism. For our numerical calculations
implemented in section\ref{IV}, substitution of representative values of 
the fundamental frequency of the pipe and the end correction of the open
mouth, i.e., $\omega \sim 2 \pi \times 913.1{\rm rad/s}$ and $\Delta L=0.005{\rm
m}$, gives $\omega \Delta L=9.131\pi \sim 28.7{\rm m/s}$. Thus the volume-flow
mechanism dominates if $V < 28.7 {\rm m/s}$. 
For the third harmonic at $2739.3{\rm Hz}$, the volume-flow mechanism
governs the driving process in the range $V< 86.1{\rm m/s}$.
As shown later, the fundamental dominates for $V \le 22 {\rm m/s}$,
while the third harmonic overcomes the fundamental for $V \ge 24{\rm m/s}$, 
then it is expected that the dominant mechanism exciting the sound wave 
in our numerical range ($2 \le V \le 40 {\rm m/s}$) is the volume flow drive.
   
\begin{figure}[bht]
\begin{center}
%\scalebox{0.8}{\includegraphics{jet.eps}}
\includegraphics[scale=0.6]{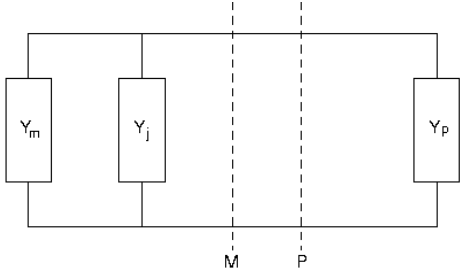}
\end{center}
\caption{Equivalent circuit network}
\label{fig-circuit}
\end{figure}

\subsubsection{Regenerative excitation mechanism} \label{IIC2}
Based on the analysis in the previous subsection, we can describe the
sound excitation mechanism of air-reed instruments. To do this, we separate the
system into a generator and a resonator by the plane, such as P or
M. Make use of an equivalent circuit network (see
Fig.\ref{fig-circuit}), we can determine the oscillating condition giving 
sound frequencies of the modes as functions of the jet velocity.  

The stability condition of the network is given by
\begin{eqnarray}
\label{eq:3Y}
Y_j+Y_m+Y_p=0,
\end{eqnarray}
where $Y_j$ denotes the jet admittance obtained from the jet motion in
eq.(\ref{eq:Jx}), % after some calculations, 
which is given by 
\begin{eqnarray}
\label{eq:Yj}
Y_j\sim \frac{VW}{\rho\omega^2\Delta L}\cosh{\mu l}
\exp\Biggl[ -i \Bigl(\frac{\omega l}{u}+\phi\Bigr)\Biggr]
\end{eqnarray}
with $\phi = \arctan\Bigl(\frac{V}{\omega \Delta L}\Bigr)$, 
the mouth admittance $Y_m$ is defined by 
$Y_m=1/Z_m=-i\frac{S_p}{\rho \omega \Delta L}$,
and the pipe admittance $Y_p$ has the forms: $Y_p=-i\frac{S_p}{\rho c}\cot kL$
for an open end pipe with a length $L$ 
and $Y_p=i\frac{S_p}{\rho c}\tan kL$ for a closed end pipe.

It is convenient to separate the condition (\ref{eq:3Y})  into the real
and imaginary parts
\begin{eqnarray}
\label{eq:ReYj}
{\rm Re}Y_j<0,\\
\label{eq:Im3Y}
{\rm Im}(Y_j+Y_m+Y_p)=0.
\end{eqnarray}
Eq.(\ref{eq:ReYj}) indicates that the jet must have a negative
resistance as a power supplier that excites the system overcoming 
non-zero resistances of the mouth and pipe impedance, which are,
however, ignored in this analysis.  
For the closed end pipe, the imaginary part (\ref{eq:Im3Y}) is rewritten as
\begin{eqnarray}
\label{eq:Im3Ycl}
&&-\frac{1}{\omega \Delta L}-\frac{V}{\omega^2\Delta L h}\cosh{\mu l}
~\sin\Biggl[ \Bigl(\frac{\omega l}{u}+
\arctan\Bigl(\frac{V}{\omega \Delta L} \Bigr) \Bigr) \Biggr]\nonumber\\
&&~~~~~~~~~~~~~~~~~~~~~~~~~~~~~~~~~~~~
+\frac{1}{c} \tan kL=0.
\end{eqnarray}
The ideal excitation is given by ${\rm Re}Y_j<0$ and ${\rm Im}Y_j=0$ so
that the phase of $Y_j$ in eq.(\ref{eq:Yj}) takes $-\pi$, i.e., $\omega l/u
=\pi-\phi$ and ${\rm Im}(Y_m + Y_p)=0$. Since ${\rm Im}Y_m<0$, ${\rm Im}
Y_p$ must takes a positive value. If the condition $\omega \Delta L/c
\ll 1$ is satisfied, we obtain $kL \sim n\pi$ for the open end pipe and
$kL\sim (n+1/2)\pi$ for the closed end pipe, which give the resonance
conditions for the open and closed end pipes, respectively.  

\begin{figure}[bht]
\begin{center}
%\scalebox{0.8}{\includegraphics{jet.eps}}
\includegraphics[scale=0.17]{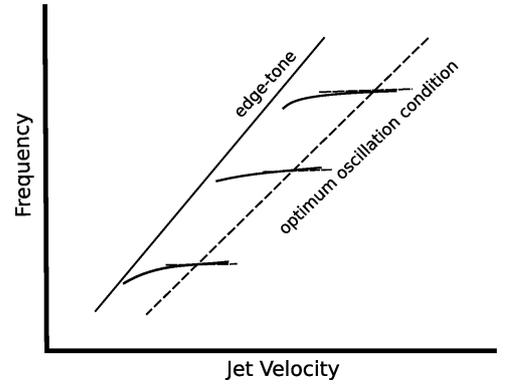}
\end{center}
\caption{Oscillation frequency vs. jet velocity}
\label{fig-coltman}
\end{figure}

Fig.\ref{fig-coltman} illustrates the frequency of excited sound wave as a
function of the jet velocity $V$  given by eq.(\ref{eq:Im3Ycl}) for the
closed end pipe together with the ideal condition $\omega l/u
=\pi-\phi$ and the edge tone frequency in eq.(\ref{eq:edge}) with $j=1$.  
The frequency of sound wave first increases in proportion to the jet velocity
similar to the edge tone, but it is synchronized with the 
fundamental resonance, if the edge tone frequency comes near the
fundamental frequency. Locking to 
the fundamental is continued until the edge tone frequency is close to
the next resonance frequency. After that it jumps up to the next
resonance and is synchronized with it, 
so the same process may be repeated every when the edge tone frequency 
reaches next one. 
%$|\phi|\ll1$
%$u \sim V/2$
%$\lambda/2 \sim l$
%$\pi/l \sim k \sim \mu$
%${\rm Im}Y_j=0$
%${\rm Im}(Y_m + Y_p)=0$
%${\rm Im}Y_j=0$

\begin{figure*}[bht]
\begin{center}
%\scalebox{0.8}{\includegraphics{jet.eps}}
(a)
\includegraphics[scale=0.24]{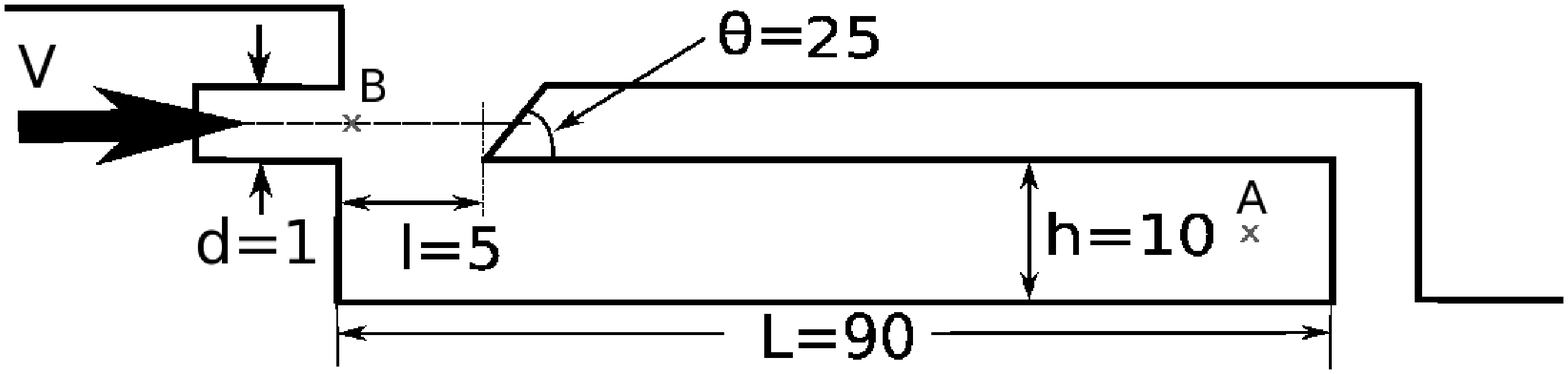}
~~~~(b)
\includegraphics[scale=0.36]{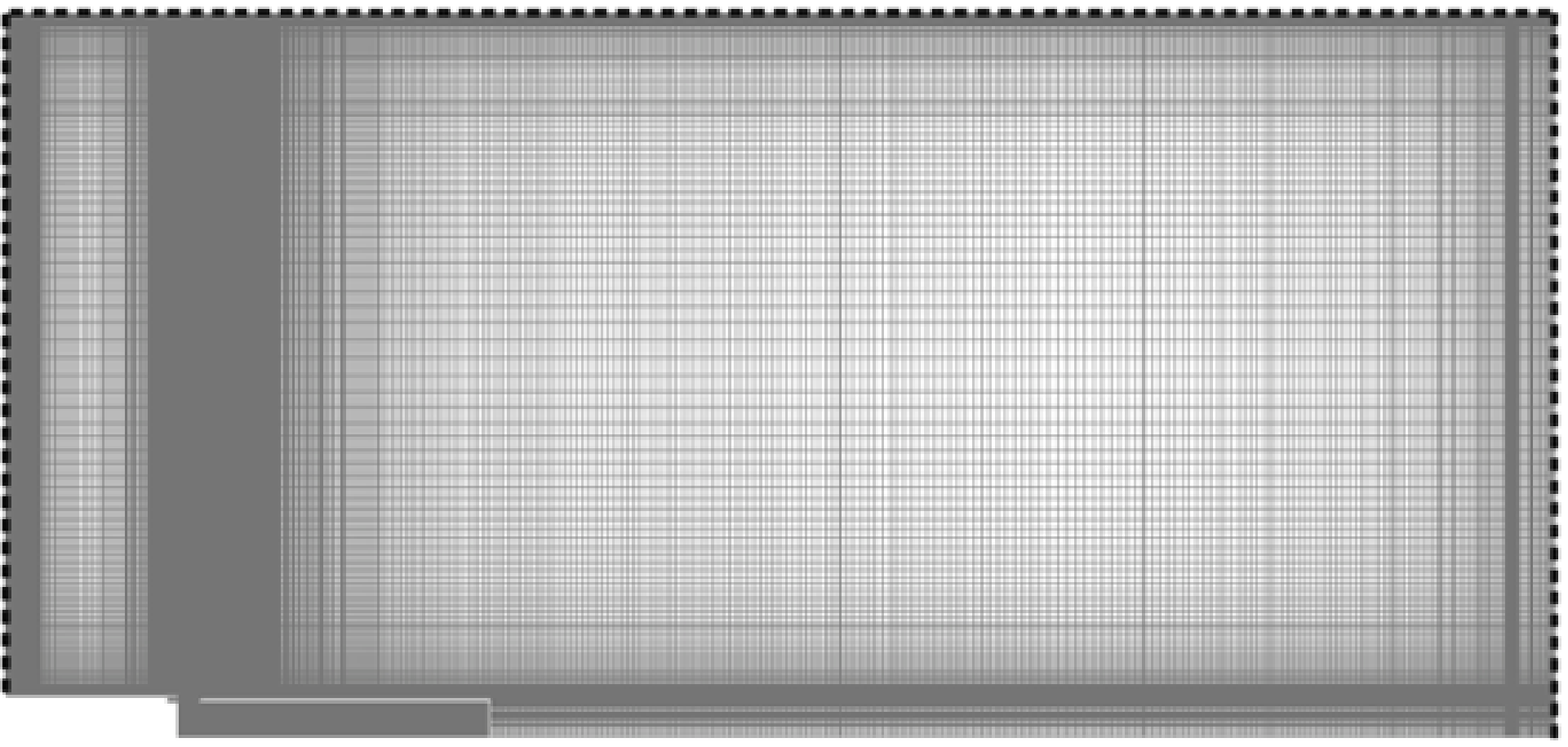}
\end{center}
\caption{(color online) Model and mesh.~
(a) Dimensions of 2D model (Unit of length :mm; Unit of angle:degree).
 (b) Numerical mesh. Dotted lines denote transparent walls.}
\label{fig-model}
\end{figure*}

\section{Model and numerical scheme}
\subsection{Numerical model}
For numerical analysis of air-reed instruments, we
simultaneously calculate the dynamics of the jet flow and the sound field
excited in a resonator.  The sound speed $c$ of about $340{\rm m/s}$ is 
much higher than the jet velocity $V$, 
which is at most several tens in MKS units. %${\rm m/s}$. 
For reproduction of sound, an extremely smaller time step is required 
compared with ordinary numerical calculations of fluid dynamics.
On the other hand, spatial scales used for calculations of fluid
dynamics with those vortices, some of which may be smaller than $1{\rm mm}$,  
are much less than wave lengths of sound, 
e.g., $34 {\rm mm}$ even at $10{\rm kHz}$. 
Therefore, in the numerical calculation of air-reed 
instruments we must satisfy the both requirements,
a sufficiently small time step to describe sound propagation and a spatial
mesh finer enough to reproduce vortices in fluid.    
Further, particle velocities of sound (or energies of sound) are 
usually much less than those of the flow. It turns out that 
sound energies in living environment are $10^{-4}$ times as small as or 
smaller than those of fluid. 
%by 4 or 5 figures. 
It is not easy to numerically calculate with a high degree of accuracy 
sound propagation dissipating for a long distance.  

To realize the calculation, we take a two dimensional (2D) model of 
a small air-reed instrument and concentrate our attention on 
dynamics in a near field of it.   
Taking a 2D model rather than a 3D model makes the number of grids
necessary for the calculation reduce, which gives rise to diminishing 
elapse time, although some of important 3D effects might be lost.   

Fig.\ref{fig-model} (a) shows geometry of the 2D model we adopt, 
where $d$ denotes nozzle height, $l$ width of mouth aperture, $L$ pipe length, 
$h$ pipe height and  $\theta$ edge angle, 
and those values are taken as shown in the figure. 
The model instrument is $9{\rm mm}$ in length thereby the fundamental
frequency being estimated as $f_0 \sim 913.1{\rm Hz}$ including the open
end correction. 
The edge angle is fixed at $25^\circ$, at which we got the most stable
oscillation in preliminary calculations. This angle value is in a suitable
range for real instruments. 

\begin{table}
\caption{Parameters of mesh}
\label{tab.1}
\begin{center}
\begin{tabular}{|c|c|c|}
\hline
points & cells & faces \\
\hline 
158,762 &  78,492  & 314,856 \\
\hline
\end{tabular}
\end{center}
\end{table}

\subsection{Numerical method}
For numerical calculation, we use a compressible
LES (Large eddy simulation), which is very popular in numerical
simulations of aero-acoustics. Actually the scheme we adopt is 
the compressible LES solver, Coodles of OpenFOAM, provided 
as a free software by OpenCFD Ltd\cite{openfoam}.
LES is very stable for a long time simulation, while it involves some
ambiguities in boundary layers 
due to statistical assumption for dynamics of eddies smaller than 
a given grid size. It however makes sense in the statistical point of view.

As shown in Fig.\ref{fig-model}(b),  
the spatial area of the numerical mesh is taken large enough 
to include an acoustic field 
outside the instrument and parameters of the mesh are taken 
as shown in Table 1.   
The averages of pressure and temperature are taken 
as $p_0=100{\rm kPa}$ and $T_0=300{\rm K}$, respectively. 
The time step of numerical integration is $\Delta t = 10^{-7}{\rm
sec}$ and time evolution up to $0.1{\rm sec}$ is obtained. 
The velocity of the jet emanating from the flue $V$ is changed 
as a control parameter in the range $(2 \le V \le 40{\rm m/s})$. 
Observations are done as follows. 
The acoustic pressure $p$ and air density $\rho$ are
observed at the point A, distance of $10{\rm mm}$ 
from the right end of the pipe and on the center line of it.  
The vorticity $\omega$ in the jet flow is
calculated at point B, distance of $1.6{\rm mm}$ right from 
the exit of the flue and on the center line of it.
%The frequencies of the acoustic pressure and
%vorticity are determined by  their  frequency spectra, respectively.

\section{Numerical results} \label{IV}
\subsection{Stable oscillation}
\begin{figure}[bht]
\begin{center}
%\scalebox{0.8}{\includegraphics{jet.eps}}
(a)
\includegraphics[scale=0.20]{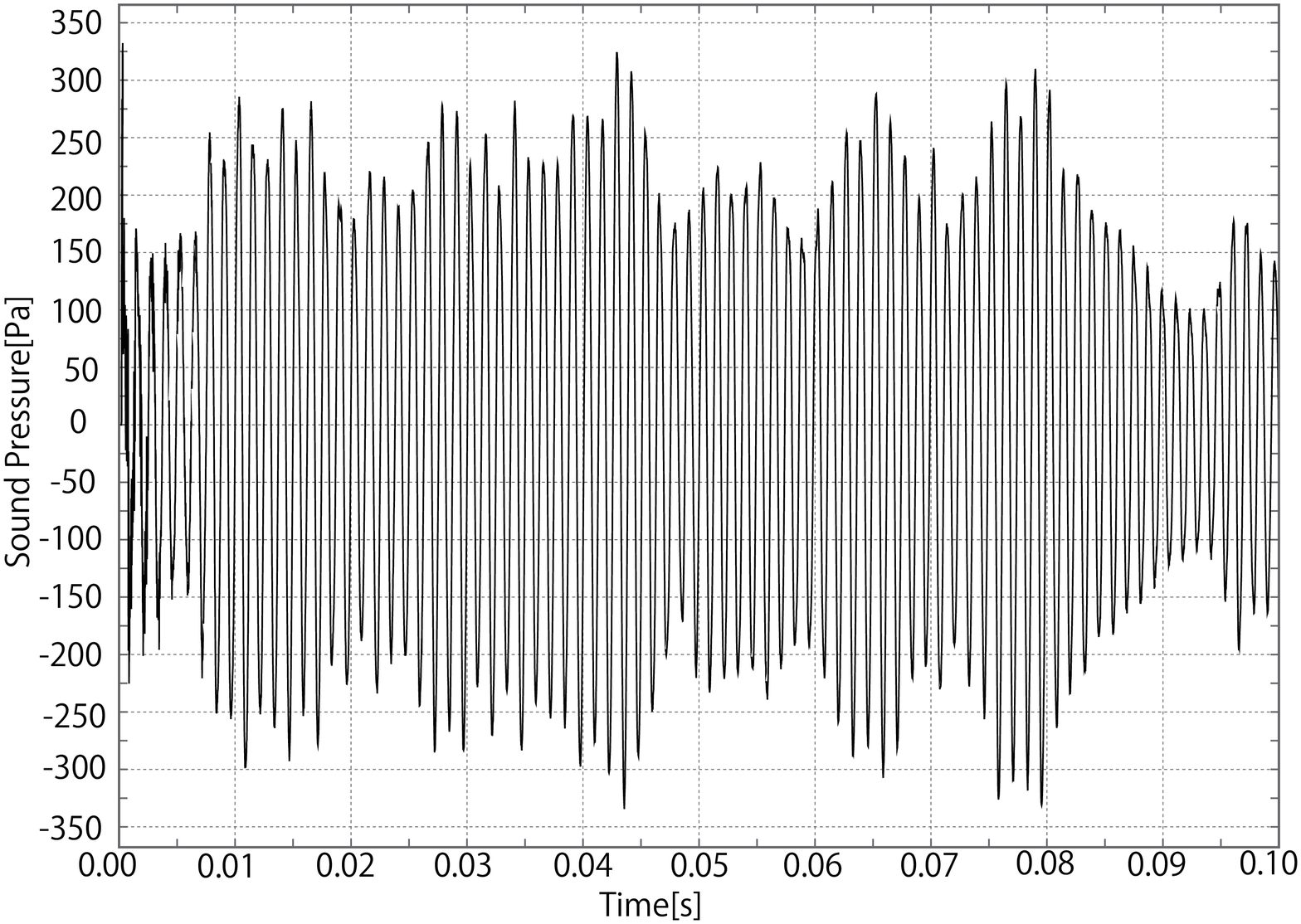}\\
\vspace{3mm}
(b)
\includegraphics[scale=0.23]{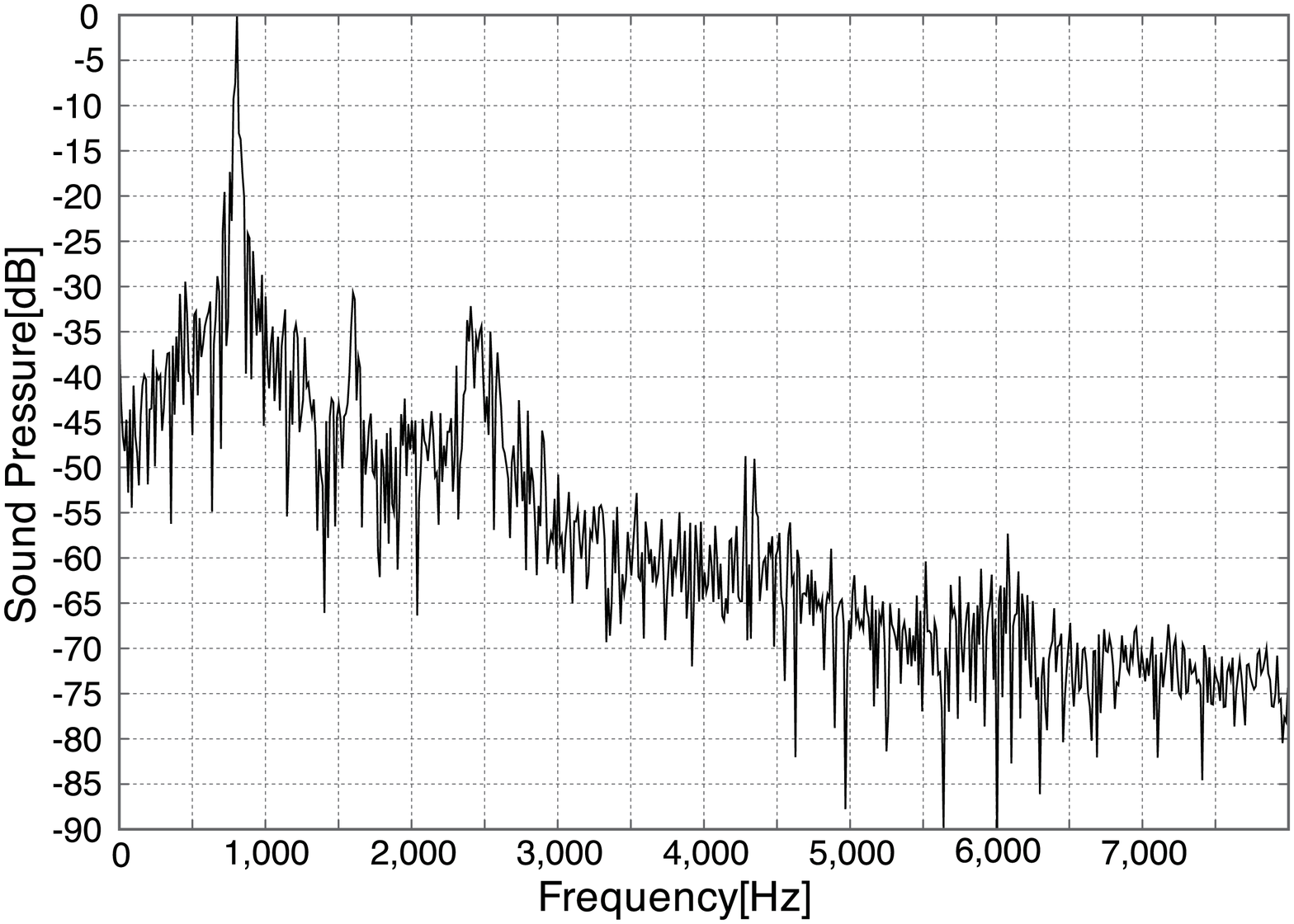}
\end{center}
\caption{Acoustic pressure at point A.~(a) Acoustic pressure oscillation.~(b) 
Power spectrum.}
\label{fig-pressure}
\end{figure}

In this subsection, we show numerical results of that oscillation at
$V=12{\rm m/s}$ which is the most stable among oscillations observed  
in the whole range of $V$.
Fig.\ref{fig-pressure} (a) and (b) show the acoustic pressure oscillation 
observed at the point A and its power spectrum, respectively. For
the calculation of the spectrum, initial transient oscillations ($0\le t
<0.01{\rm s}$) are omitted. 
The oscillation of the acoustic pressure in 
Fig.\ref{fig-pressure} (a) is very stable with an almost constant pitch 
in the whole range of time ($0\le t \le 0.1{\rm s}$) 
except for a short initial transition.  
Its amplitude is gently undulated but reaches up several hundreds ${\rm
Pa}$, which are much larger than normal acoustic pressures in the open 
air field.  
As shown in Fig.\ref{fig-pressure}(b), the main peak of the spectrum is very
sharp and appears at $\nu=805.7{\rm Hz}$, which is less than the theoretical
estimation of the fundamental resonance frequency $913.1{\rm Hz}$ 
by $107.4{\rm Hz}$. However, it rapidly approaches the resonance
frequency with increase of $V$ as shown later.
Thus it is considered that the oscillation at $V=12{\rm m/s}$ is almost 
locked on the fundamental resonance.      

\begin{figure}[bht]
\begin{center}
%\scalebox{0.8}{\includegraphics{jet.eps}}
(a)
\includegraphics[scale=0.235]{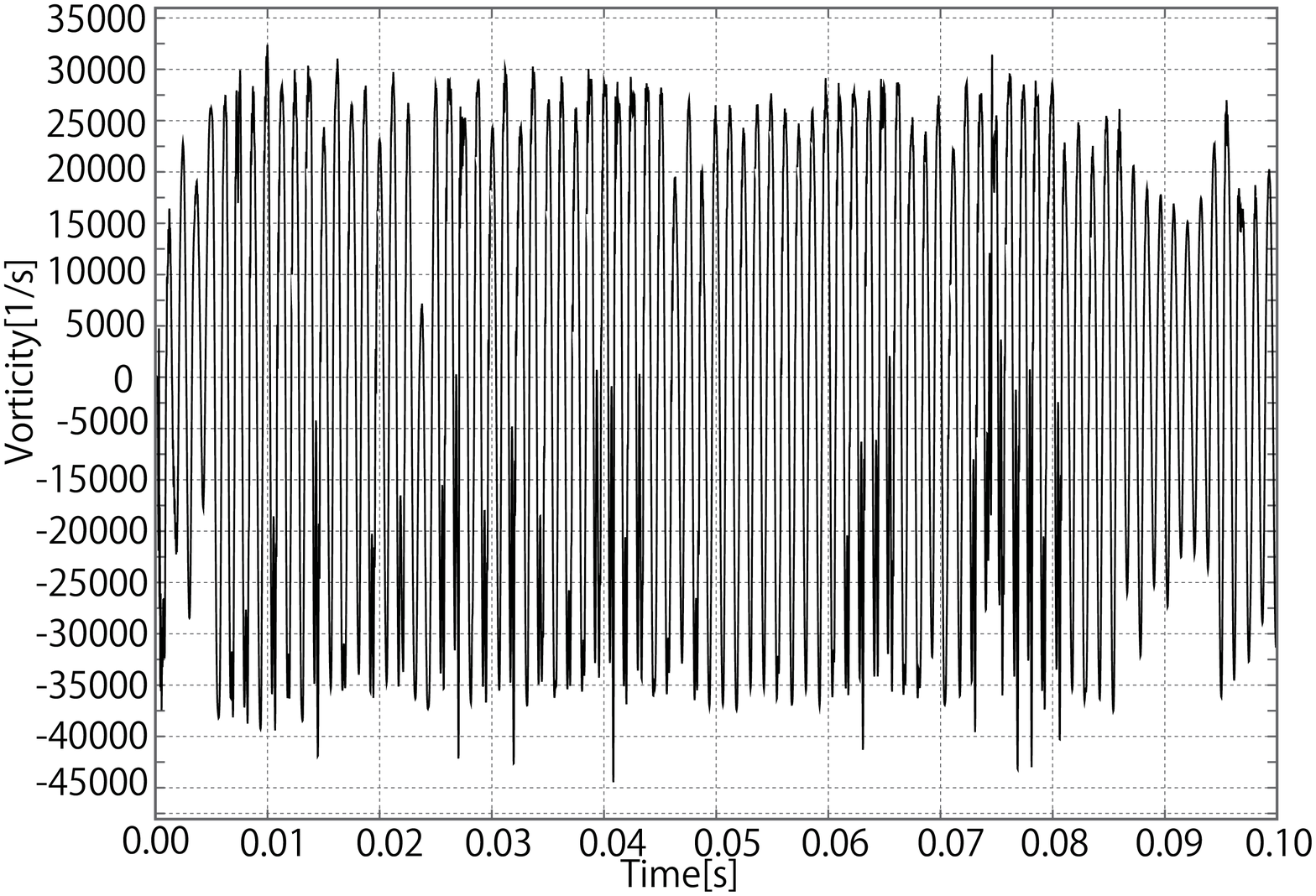}\\
\vspace{3mm}
(b)
\includegraphics[scale=0.23]{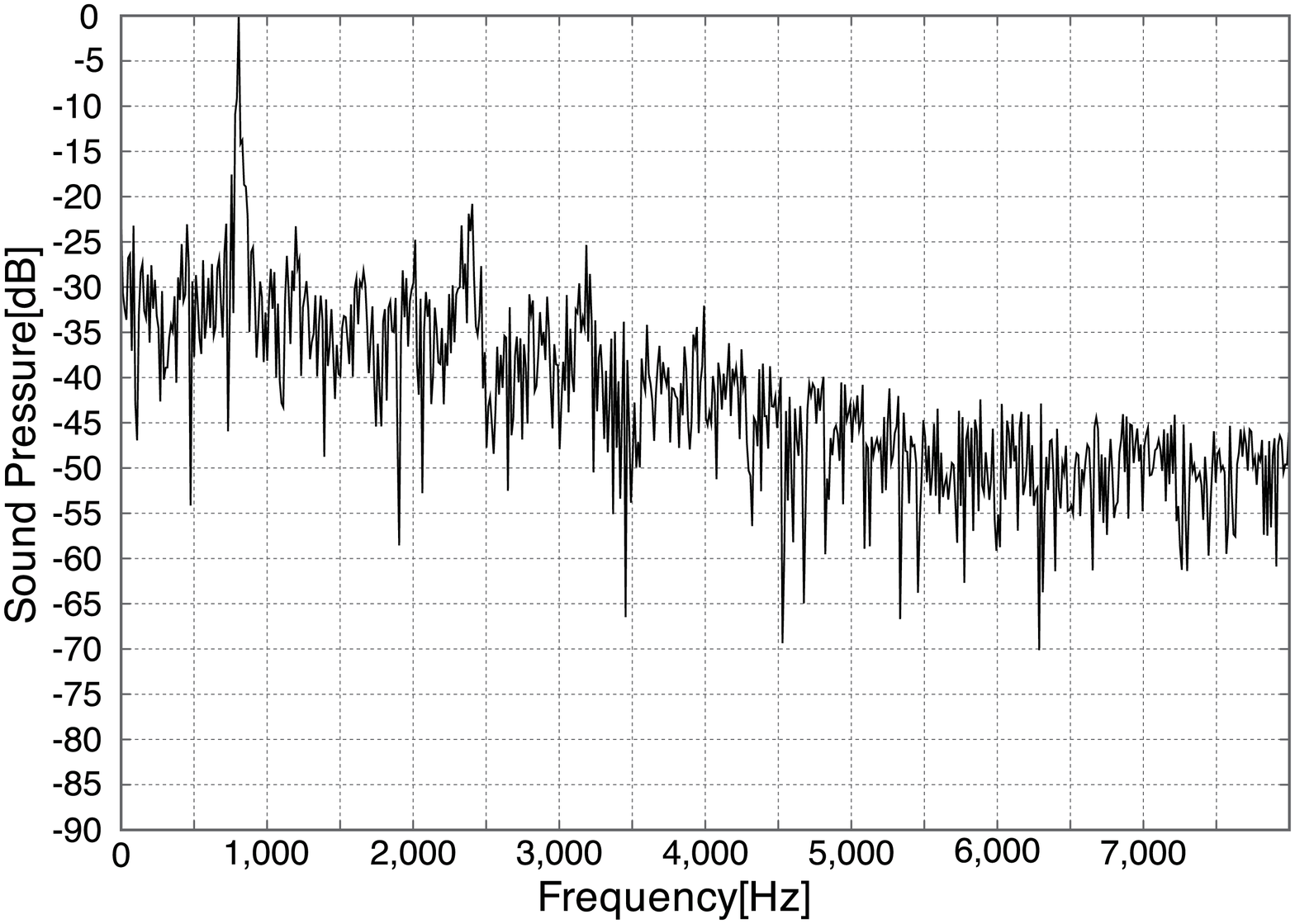}
\end{center}
\caption{Vorticity at point B.~
(a) Vorticity oscillation.~(b) Power spectrum.}
\label{fig-jetvor}
\end{figure}

\begin{figure}[bht]
\begin{center}
%\scalebox{0.8}{\includegraphics{jet.eps}}
\includegraphics[scale=0.23]{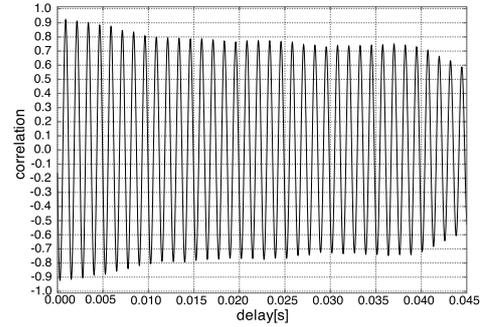}
\end{center}
\caption{Correlation between sound pressure and jet vorticity.}
\label{fig-soukan}
\end{figure}

\begin{figure*}[bht]
\begin{center}
%\scalebox{0.8}{\includegraphics{jet.eps}}
(a)
\includegraphics[scale=0.33]{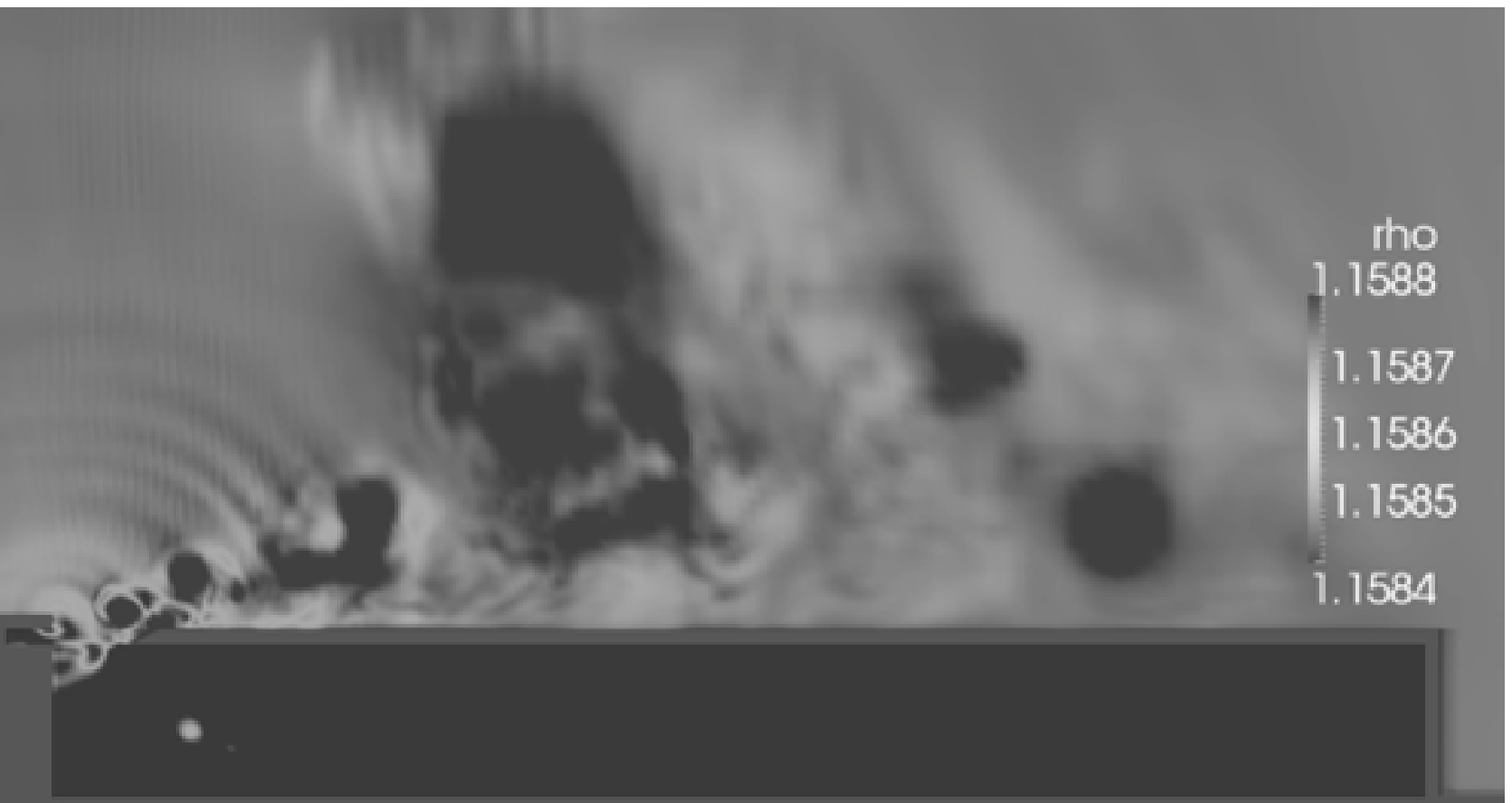}
~~~~~(b)
\includegraphics[scale=0.33]{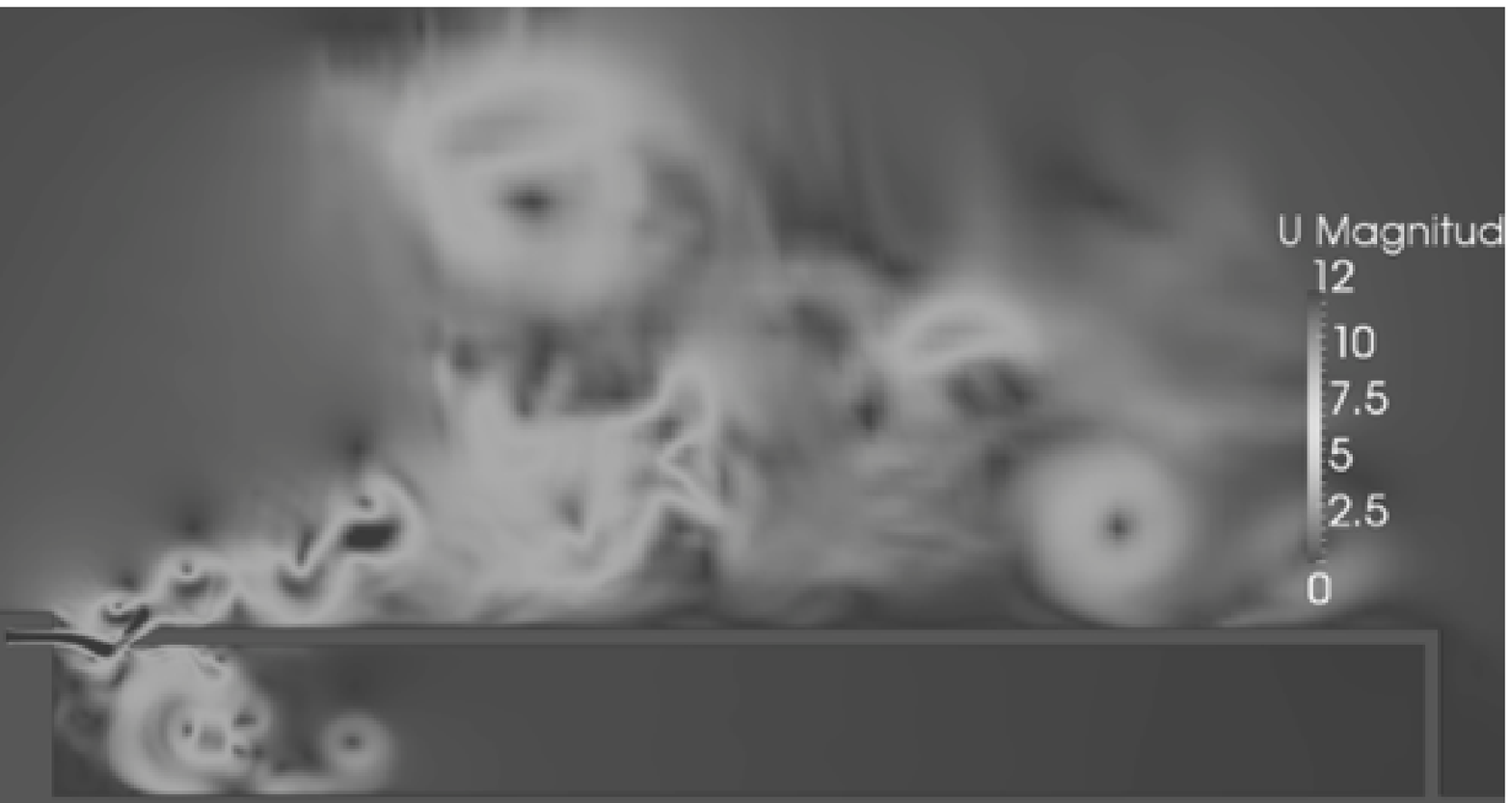}\\
\vspace{5mm}
(c)
\includegraphics[scale=0.33]{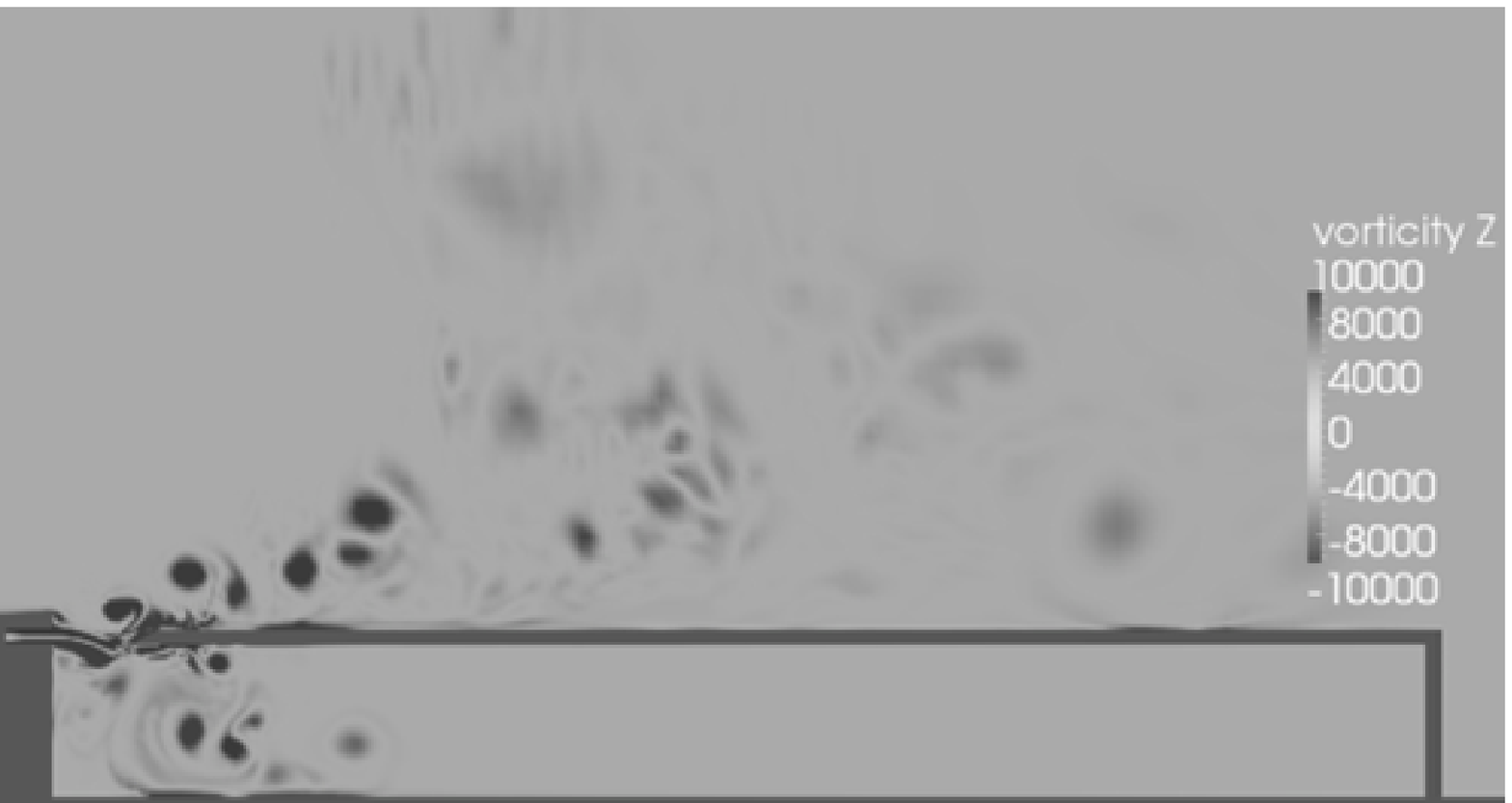}
~~~~~(d)
\includegraphics[scale=0.33]{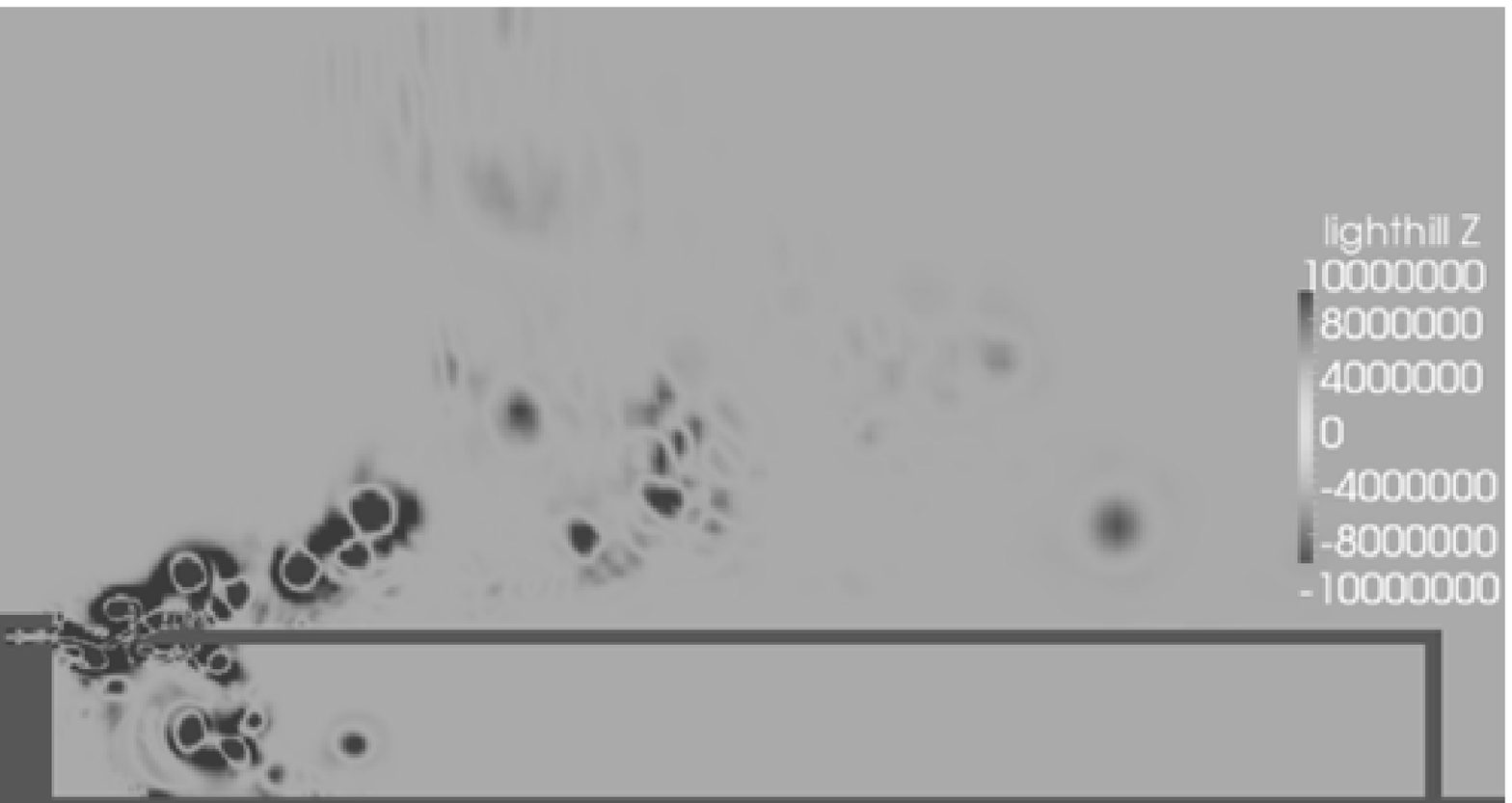}
\end{center}
\caption{(color online) Spatial distributions of representative dynamical variables.~
(a) Air density.~(b) Flow velocity.~
(c) Vorticity.~(d) Lighthill's sound source.}
\label{fig-distall}
\end{figure*}

Fig.\ref{fig-jetvor} (a) and (b) show the vorticity fluctuation of jet 
observed at the point B and its power spectrum, 
respectively. The vorticity regularly oscillates like the acoustic
pressure at the point A with the same fundamental frequency as shown 
in the power spectrum(see Fig.\ref{fig-jetvor} (b)). 
Fig.\ref{fig-soukan} shows the mutual correlation function between the 
acoustic pressure ($0.01\le t \le 0.1{\rm s}$) and the vorticity
($0.01\le t \le 0.1{\rm s}$).   
The correlation changes periodically with
the same fundamental frequency and does not decay for a long time. It
means that there is a strong correlation between the jet motion and sound
field in the resonator and so the jet motion is
controlled rather by the pipe resonance than by its self-oscillation 
mechanism.

Fig.\ref{fig-distall} shows spacial distributions of characteristic dynamical
variables at a certain time: air density, flow velocity,
vorticity and Lighthill's sound source.  
We used eq.(\ref{eq:2DLighthill}) for the calculation of 
Lighthill's sound
source. Though it is not shown in a picture, the acoustic pressure
distribution $p(x,t)$ is almost as same as the air density 
distribution $\rho(x,t)$. In the stationary oscillation regime, 
fluctuations of $\rho(x,t)$, more precisely $|\rho(x,t)-\rho_0|$, 
become much larger in the inside of pipe than in the outside as shown in
Fig.\ref{fig-distall} (a).      
 
In the velocity distribution in Fig.\ref{fig-distall}(b), 
we see that eddies mainly arise from
collision of the oscillating jet with the edge and are pouring into the
inside or going to the outside of pipe. 
The eddies in the outside run along the wall of pipe being 
dissipated gradually. 
On the other hand, those in the inside roll up and make 
a large rotor or a few rotors near the mouth opening not spreading into more
right hand side.  No eddies apparently appear in the right 3/4 part of the
pipe, in which the strong sound field dominates the flow. 
Such a rotor(s) always appears in stable oscillations and so 
the existence of the rotor(s) may affect the jet motion, 
though we do not touch details of the mechanism 
in this paper. 

As shown in Fig.\ref{fig-distall}(c), vorticity takes large (absolute) 
values along the upper and lower boundaries of jet and in areas of 
strong eddies rolled up.  Lighthill's sound source in
Fig.\ref{fig-distall}(d)  almost overlaps with the vorticity
distribution. This fact is supported by the main claim of the 
Powell-Howe vortex sound theory\cite{Howe2, Powell,Howe} 
that the major part of Lighthill's sound sources is
of contribution of vortices. The sound sources along the jet
periodically oscillate reflecting the motion of jet. 
On the other hand, the other 
sound sources mainly located on large eddies behave rather
irregularly in time evolution due to irregular motions of the eddies.     
This fact combined with the fact that the vorticity of jet is
strongly correlated with the sound field excited in the pipe 
suggests that the main mechanism driving the sound field in the pipe at
$V=12{\rm m/s}$ is 'volume flow drive' rather than 'momentum drive'. 
This suggestion is also supported by the theoretical prediction
discussed in subsection \ref{IIC1}.     

\subsection{Frequency change with jet velocity}  
\begin{figure}[bht]
\begin{center}
%\scalebox{0.8}{\includegraphics{jet.eps}}
\includegraphics[scale=0.25]{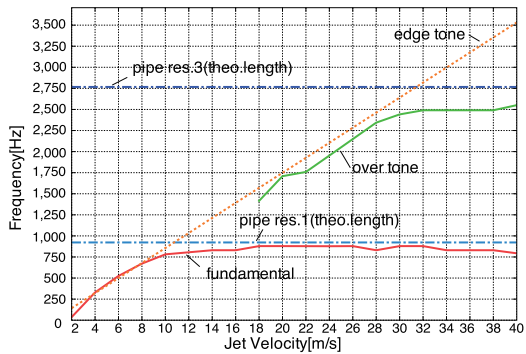}
\end{center}
\caption{(color online) Changes of oscillation frequencies with jet velocity. The lines
labeled 'pipe res.1' and 'pipe res.3' respectively indicate the frequencies of
 fundamental and 3rd harmonics resonances estimated theoretically.}
\label{fig-fvsV}
\end{figure}

In this subsection, we discuss change of characteristic frequencies of 
sound waves excited in the pipe, i.e., spectrum peaks of fundamental and first 
overtone, with increase of the jet velocity $V$. 
In Fig.\ref{fig-fvsV}, we show characteristic frequencies of the acoustic 
pressure at the point A %together with those of the jet vorticity 
%at the point B 
as functions of V in the range 
($2\le V \le 40{\rm m/s}$). For comparison, resonance frequencies of 
the pipe estimated theoretically and the edge tone frequency given 
by eq.(\ref{eq:edge}) with $j=1$ are also depicted in Fig.\ref{fig-fvsV}. 
%The characteristic frequencies of the acoustic pressure almost coincide 
%with those of the jet vorticity in the whole range of V.  
%It means that the motion of the jet is synchronized with the 
%sound wave and vice versa.
The fundamental frequency of the acoustic pressure %or the vorticity 
is observed in the whole range of $V$, but the first overtone is
identified only at high velocities, $V\ge 18{\rm m/s}$.

\begin{figure}
\begin{center}
%\scalebox{0.8}{\includegraphics{jet.eps}}
(a)
\includegraphics[scale=0.20]{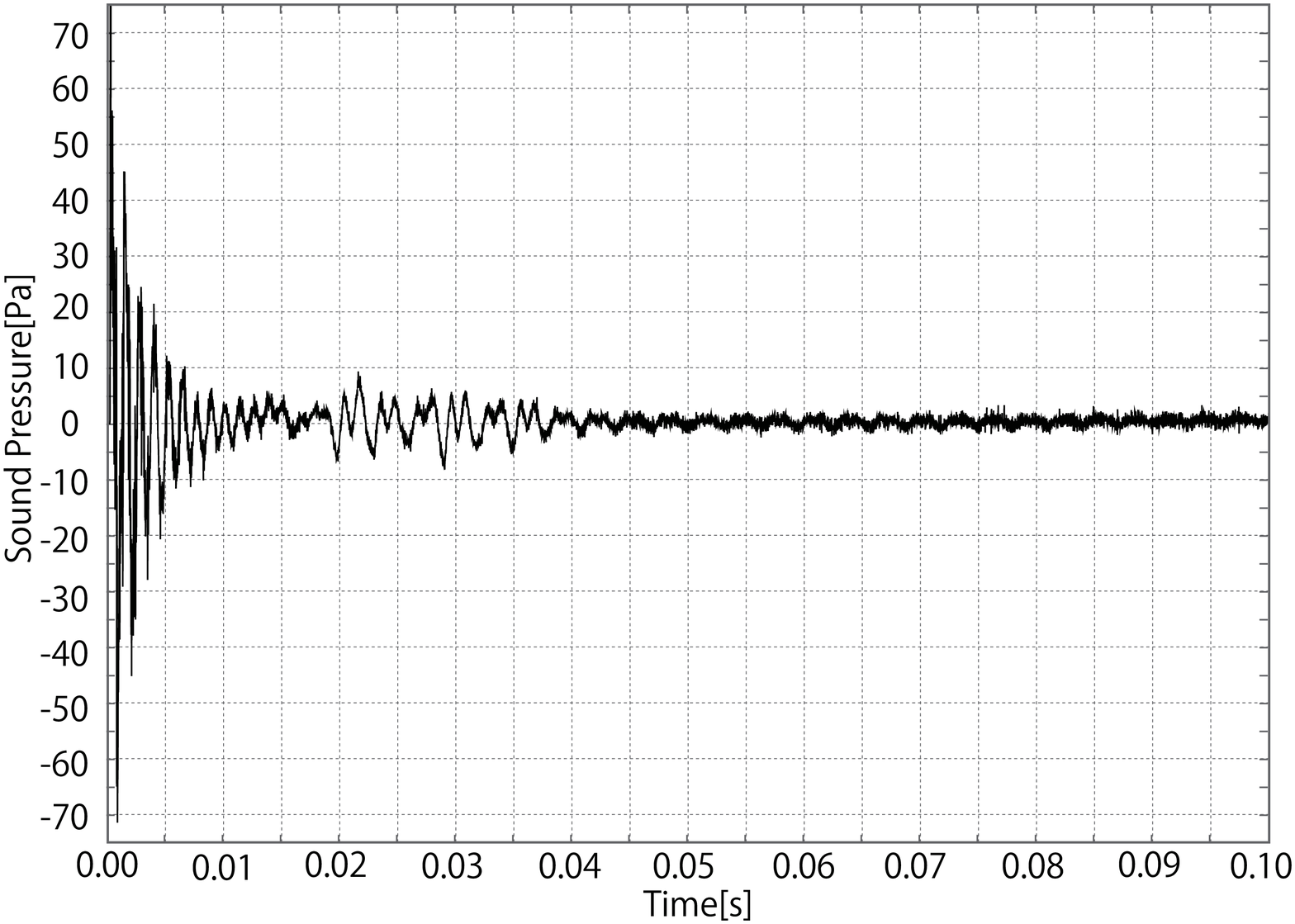}\\
(b)
\includegraphics[scale=0.23]{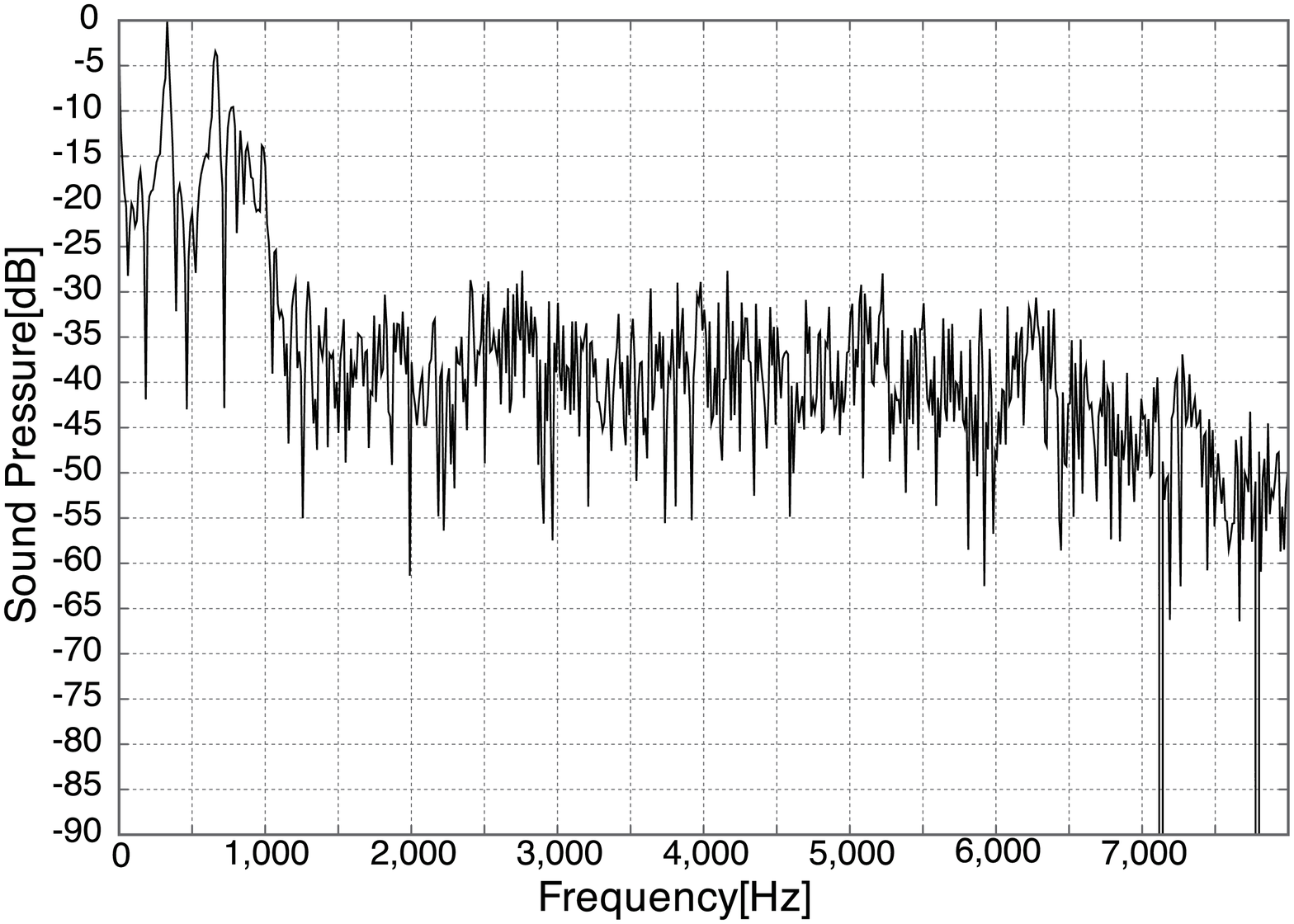}\\
(c)
\includegraphics[scale=0.23]{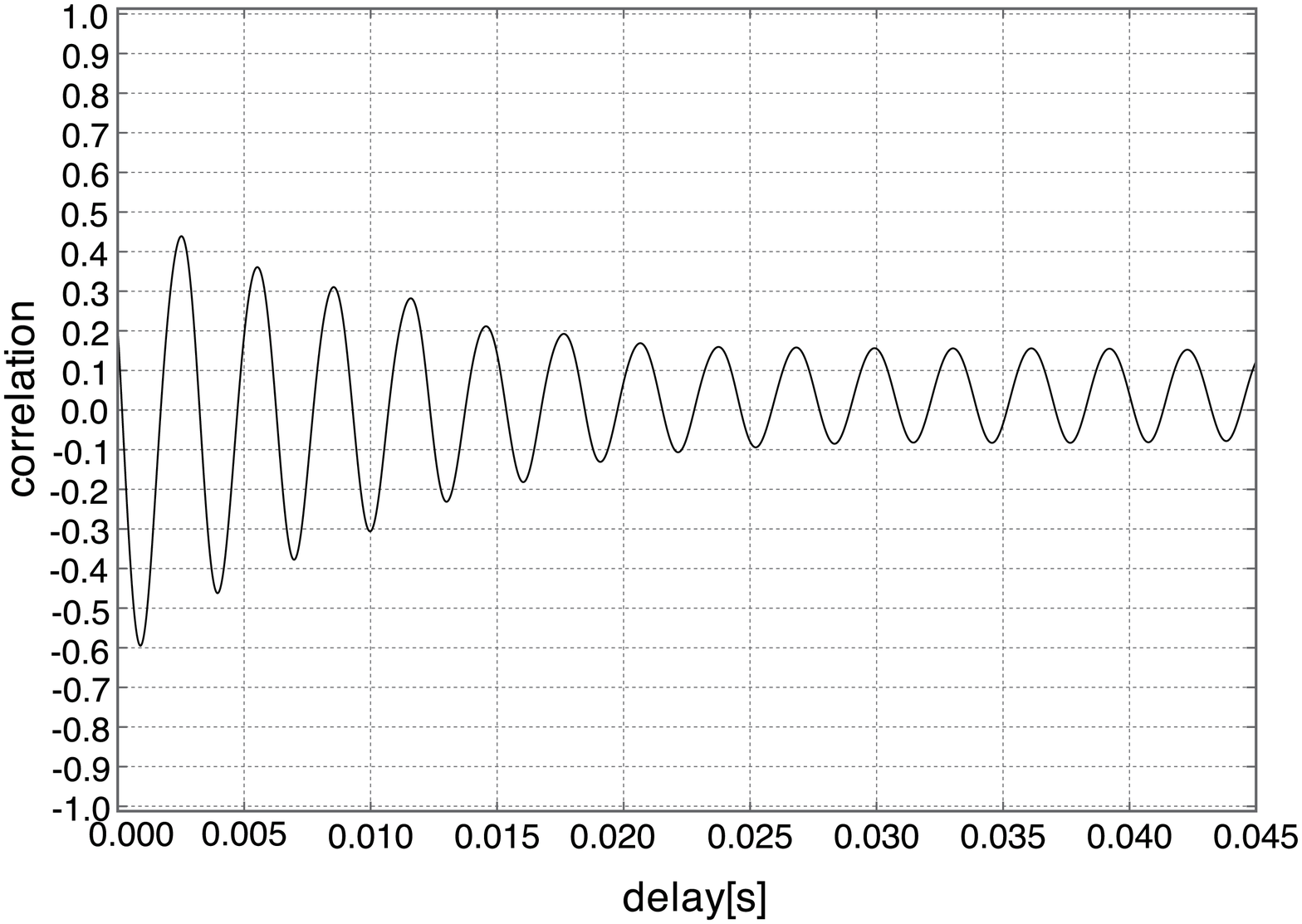}
\end{center}
\caption{Oscillation at $V=4{\rm m/s}$.
~(a) Acoustic pressure.~(b) Power spectrum. (c) Correlation with vorticity at point B.}
\label{fig-pressure4}
\end{figure}

In the low velocity regime, ($V \le 8{\rm m/s}$), the fundamental 
frequency of acoustic pressure increases as that of the edge tone, namely 
it is proportional to $V$. 
Hence the jet motion is little affected by the pipe resonance  
and almost keeps its natural oscillation. 
For example, the acoustic pressure and its power spectrum at 
$V=4{\rm m/s}$ are shown in Fig.\ref{fig-pressure4} (a) and (b), respectively. 
The acoustic pressure oscillates somewhat regularly at 
$\nu_0=341.8{\rm Hz}$ with very small amplitudes compared with 
those at $V=12{\rm m/s}$.    
However, the power spectrum shown in Fig.\ref{fig-pressure4}(b) is something 
interesting: the first overtone peak at 
$\nu_1=659.2{\rm Hz}$ accompanied by smaller side peaks in a upper range 
is observed. 
This broad band of peaks seems to be affected by the pipe resonance. 
Indeed, the fundamental
frequency $\nu_0$ is slightly less than half of fundamental
resonance frequency, so  a broad band of
peaks accompanying the overtone can appear near the fundamental 
resonance frequency.   
As shown in Fig.\ref{fig-pressure4} (c), the correlation between 
the acoustic pressure and vorticity oscillates in a regular manner but  
gradually decays to an oscillation much smaller than that at 
$V=12{\rm m/s}$. Then it is considered
that the interaction between the generator and resonator is small   
in this regime. Namely an unidirectional interaction from the generator 
to the resonator dominates the operation with less feedback from the
resonator, thereby being out of resonance. 

In the middle range, ($10\le V \le 22{\rm m/s}$), oscillations of 
sound wave locking on the fundamental pipe resonance are
observed. The frequency of oscillation is fairly lower than that of 
fundamental resonance at $V=10{\rm m/s}$, but it quickly approaches the 
frequency of the resonance with increase of $V$. The oscillations are 
very stable in the range ($10\le V \le 16{\rm m/s}$) and the most 
stable one is observed at $V=12{\rm m/s}$ as shown in the previous 
subsection. So the oscillation is most stabilized
just above the value of $V$ at which the jet motion starts
synchronizing with the pipe resonance.  
In the range ($18 \le V \le 22 {\rm m/s}$), oscillations, however, 
become slightly unstable. Namely some amplitude modulations occur in 
long term evolutions, though we do not show results. In the spectra of 
those oscillations, there observes a peak corresponding to the edge tone
frequency at the present velocity $V$. Further smaller peaks whose
frequencies are nearly equal to the third harmonic of the pipe are often 
observed. It is considered that 
competition among the resonances and edge tone, i.e., 
inherent oscillation of the jet, occurs, but the fundamental still 
dominates the others, though its oscillation is somewhat disturbed. 

\begin{figure}
\begin{center}
%\scalebox{0.8}{\includegraphics{jet.eps}}
(a)
\includegraphics[scale=0.20]{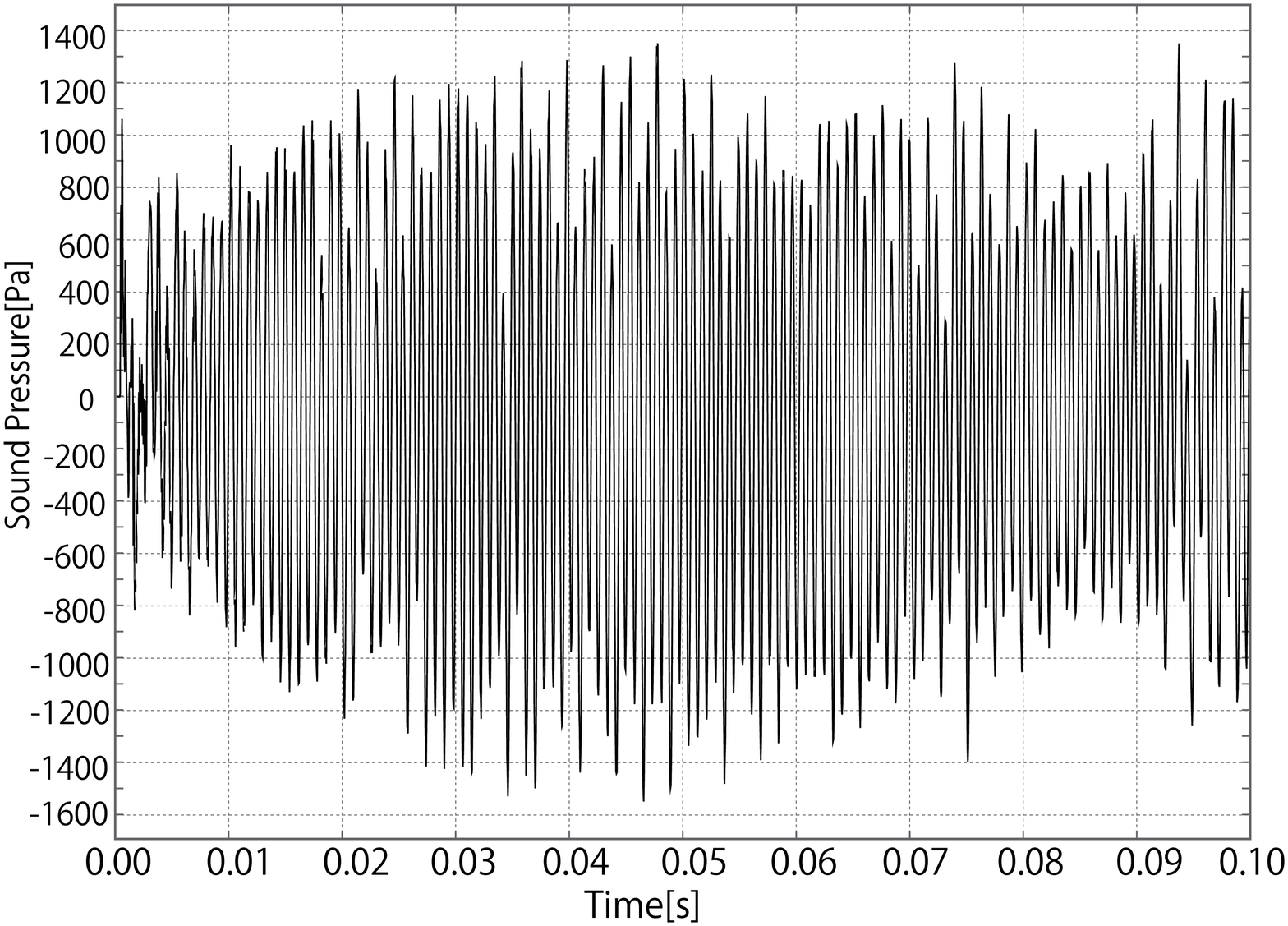}\\
(b)
\includegraphics[scale=0.23]{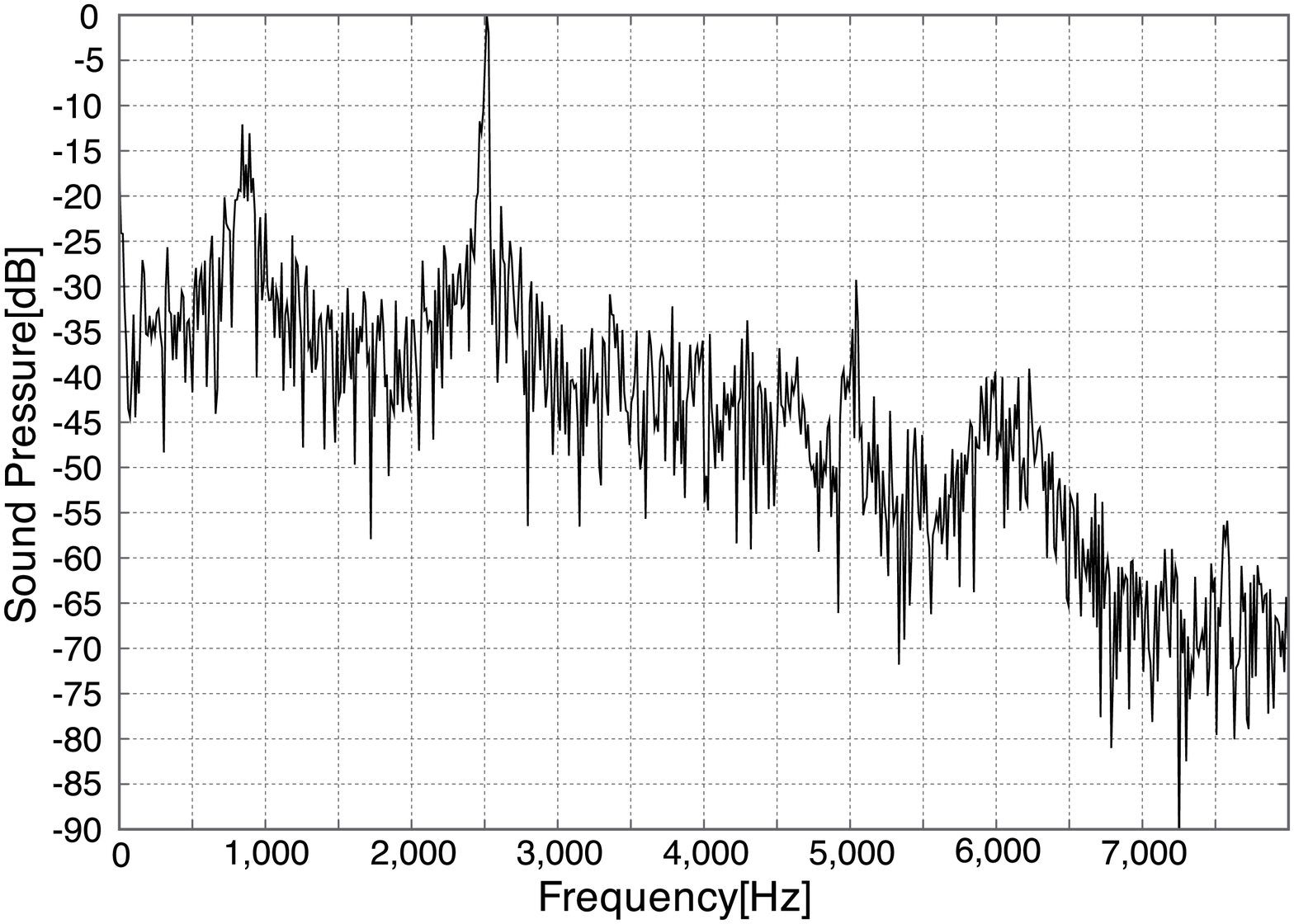}\\
(c)
\includegraphics[scale=0.23]{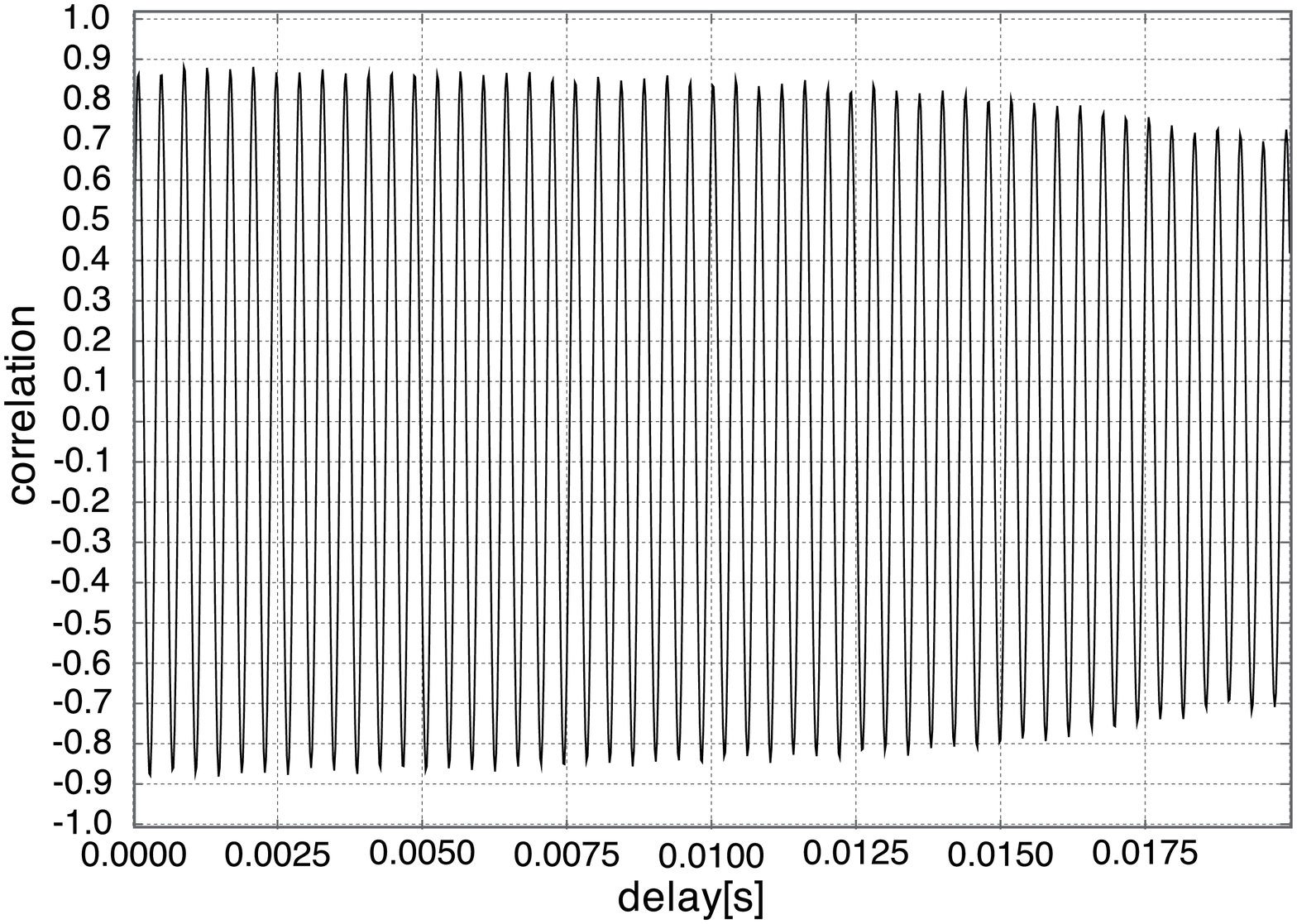}
\end{center}
\caption{Oscillation at $V=36{\rm m/s}$.~(a) Acoustic pressure.~(b) Power
spectrum. (c) Correlation with vorticity at point B.}
\label{fig-pressure36}
\end{figure}

In the high velocity range, ($V \ge 24{\rm m/s}$), first overtone peaks
likely to be the third harmonic are clearly observed in 
the spectra of acoustic pressure.  The frequency of the first overtone
peak increases mostly proportional to the jet velocity like the edge tone
in the range $18 \le V \le 24{\rm m/s}$ but almost converges to a constant
value in the range $V \ge 32 {\rm m/s}$, e.g., 
$\nu_1=2490.2{\rm Hz}$ at $V=36{\rm m/s}$, 
although it is considerably lower than the theoretical estimation of the
third harmonic, $2739.3{\rm Hz}$. 
%Unfortunately, we don't have any idea to
%explain the discrepancy between the theoretical prediction and the
%observed values.   
Then the oscillation would start to
be synchronized with the third harmonic around $V=24\sim26{\rm m/s}$.
Indeed, the peak height of the first overtone is increasing with increase of
$V$ and becomes larger than that of the fundamental for $V\ge 24{\rm
m/s}$ (for example see Fig.\ref{fig-pressure36}(b) at $V=36{\rm m/s}$). 
Further the wave form of acoustic pressure changes to that of the 
third harmonic for $V\ge 24{\rm m/s}$.  
Fig.\ref{fig-pressure36}(a) shows the time evolution of 
the acoustic pressure at $V=36.0{\rm m/s}$. 
It oscillates almost periodically with very strong
amplitudes, sometimes exceeding $1{\rm kPa}$. 
%830,1Hz
As shown in Fig.\ref{fig-pressure36}(c), the correlation between the sound 
pressure at the point A and the vorticity at the point B behaves in a
regular manner and does not decay in a long term. 
In conclusion, it can be said that the transition 
from the fundamental to the third harmonic occurs around $V=24{\rm m/s}$.

\begin{figure}[bht]
\begin{center}
%\scalebox{0.8}{\includegraphics{jet.eps}}
\includegraphics[scale=1]{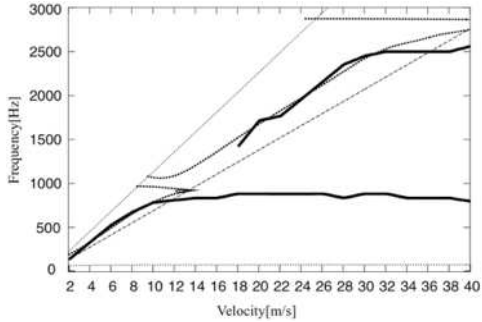}
\end{center}
\caption{Comparison of the numerical result with the theoretical
 prediction in change of frequencies with the jet velocity.}
\label{fig-fvsVth}
\end{figure}

We compare our numerical result with the theoretical
prediction introduced in section \ref{IIC2}. 
Fig.\ref{fig-fvsVth} shows the velocity-frequency curves (broken lines) 
given by eq.(\ref{eq:Im3Ycl}) at parameter values adjusted to 
our numerical calculation: $l=5{\rm mm}$, $\Delta L=9{\rm mm}$,
$h=10{\rm mm}$, $L=90{\rm mm}$, $c=340{\rm m/s}$, $\mu l=\pi$. However,
we need to rescale the argument of sinusoidal function in
eq.(\ref{eq:Im3Ycl}) %$\omega l/u$ 
as $u/l=0.8V/l$ to make it be
adjusted to the edge tone equation (\ref{eq:edge}) and also to our
numerical result, while it is usually taken as $u/l\le 0.5V/l$. 
The curves of eq.(\ref{eq:Im3Ycl}) are drawn
only in the area limited by the two dotted lines given by ${\rm Re}Y_j=0$, 
in which the jet has a negative value in resistance, the necessary condition
to excite the instrument.   
The chain line stands for the optimum oscillation condition ${\rm Im}Y_j=0$.  
For comparison, the velocity-frequency curves obtained
numerically, which are the sames as those in Fig.\ref{fig-fvsV}, 
are also drawn by solid lines.  

Our numerical result shows good agreement with the
theoretical prediction in three characteristic ranges: oscillations of
the edge tone in the low velocity range, locking to the
fundamental resonance in the middle range, and transition and locking to
the third harmonic in the high range. It is expected that the optimum
oscillations of the first and third harmonics occur at intersections
of the corresponding branches of eq.(\ref{eq:Im3Ycl}) 
with the line ${\rm Im}Y_j=0$, respectively. The intersection appears 
just after locking to the first or third harmonics starts.    
The fact that the optimum oscillation of the fundamental is observed
numerically at
$V=12{\rm m/s}$ well supports the validity of theoretical prediction. 
A well sustained third harmonic, e.g., at $V=36{\rm m/s}$, is also
observed near the intersection of the third harmonic.  

%double-dot chain line

Here, we explain why we take the parameter $u/l$  
as $u/l=0.8V/l$ instead of $u/l=0.5V/l$ in order to calculate the
velocity-frequency curves and optimum line. 
According to experiments and the 
semi-empirical theory, the velocity of jet wave $u$ should take a value in
the range $u \le 0.5V$. However, even at $u=0.5V$, the
velocity-frequency curves given by eq.(\ref{eq:Im3Ycl}) lean to the right more. 
As a result, the curves are markedly shifted from the edge
tone line in a low velocity range and the onset point of 
locking to the fundamental is estimated as $V\sim 20{\rm m/s}$. 
Then those curves apparently go out of our numerical 
results and also out of Brown's edge tone equation.  
Taking into account the fact that the existence of the 
edge is ignored in framing the jet model in that theory
(see \ref{IIC1}), it is considered that such extent of
modification is still within an acceptable limit. 
It is rather surprised that the semi-empirical
theory shows quantitatively good agreement with full numerical
calculations with such a slight modification.       

Finally note that as discussed in subsection \ref{IIC1}, 
it is predicted that the volume-flow mechanics dominates in excitation
of sound waves not only in the range of
the fundamental but also in the range of the third harmonic.
We think that this prediction is probably true and is partially supported by 
the observation that the correlation between the acoustic pressure and 
the vorticity of the jet behaves regularly without remarkable decays 
in the ranges of $V$, in which the fundamental and the third harmonic
are well sustained.  
Further it is also confirmed in preliminary calculations 
that the correlation becomes more unstable, 
if the vorticity is observed at a point in the relaxation area 
between the planes  M and P in Fig.\ref{fig-opptrecorder}, i.e.,
momentum driving area,  
because motions of eddies in this region are more irregular than the jet
oscillation.      

\section{Summary and discussion}

In this paper, we have reported on the numerical analysis of 
a 2D air-reed instrument with the compressible LES. 
As a result, vibrations of the 
air-reed instrument are well reproduced by using the compressible LES.
Especially, some characteristic features of the air reed instrument 
obtained numerically show good agreement with
those given by the theoretical prediction as well as experimental
results\cite{PhysMI,Coltman3,Coltman4,Fletcher,Elder,Hirschberg_group3}.
The characteristic features reproduced are as follows. 

When the velocity is small enough, the inherent jet oscillation 
radiating the edge tone is predominant so that the oscillation
frequency of sound excited
in the pipe is almost proportional to the jet velocity. 
However, the jet oscillation is synchronized with the fundamental resonance 
of the pipe when the frequency of edge tone approaches that of the
fundamental resonance with increase of the jet velocity.
The synchronization with the fundamental resonance continues 
until the frequency of edge tone reaches that of the first overtone, 
i.e., 3rd harmonic in our case. 
Then the synchronization is turned off and
the transition to the 3rd harmonic arises, namely the 
frequency locking to the 3rd harmonic takes place.   

Based on the results of this paper, we are able to proceed to study the
acoustic mechanism of the air-reed instrument comprehensively. 
With help of the Lighthill theory\cite{Lighthill} and/or 
Powell-Howe vortex sound theory\cite{Howe,Powell,Howe2}, 
the place of sound sources in turbulence will be detected
clearly and their behavior will be characterised in terms of
aero-dynamics and nonlinear dynamics. 
The following problems and questions should  be investigated. 
The questions which mechanism, volume-flow mechanism or momentum drive 
mechanism\cite{Cremer,Coltman,Coltman2,Coltman3,Coltman4,Fletcher,Elder,Yoshikawa}, dominates at a
given playing condition and what change occurs with change of
the playing condition, e.g., with the jet velocity, 
should be answered clearly.     
It should be also clarified what kind of influence the resonance of the pipe 
exerts on the motion of the jet, because it must be the key to understand
%different from the
%mechanism of the edge tone without a resonator and
the synchronization mechanism between the jet flow and pipe resonance, 
%is the key to understand 
the key mechanism of wind instruments different from the edge tone. 

To analyze more realistic behavior of instruments, the numerical 
analysis of 3D models is needed. In a preliminary
calculation, we have succeeded to numerically reproduce acoustic
oscillations of the ocarina with a 3D model\cite{Kobayashi}. 
It is considered that the ocarina uses the Helmholtz resonance caused by 
an elastic property of air instead of the pipe resonance.  
Then, it is quite interesting to consider the problem how different acoustic
mechanisms work for different types of resonator 
with comparing numerical data of the ocarina with that 
of a 3D model of the instrument with a resonance pipe. 
Before that it is of course necessary to clarify differences between the
2D and 3D models. 

\begin{acknowledgments}
This work is supported by Grant-in-Aid for Exploratory Research  
No.20654035 from Japan Society for the Promotion of Science (JSPA).
\end{acknowledgments}

%\bibliography{takahasi}
%\end{document}

\end{document}